\documentclass[twocolumn]{autart}    
\usepackage{color}
\usepackage{harvard}
\usepackage{graphicx}          

\usepackage{amsfonts}
\usepackage{amsmath}
\usepackage{amssymb}
\usepackage{bbm}
\usepackage{amstext,framed,pdfsync}
\usepackage{amssymb}
\usepackage{subcaption}
\usepackage{cite}

\newcommand{\tx}[1]{\text{#1}}

\newcommand{\vx}[1]{\mathbf{#1}}
\newcommand{\wf}{{}^{\Wf}}
\newcommand{\hf}{{}^{\Hf}}
\newcommand{\wfs}{_{\Wf}}
\newcommand{\hfs}{_{\Hf}}
\newcommand{\Wf}{\tiny{\mathcal{W}}}
\newcommand{\Hf}{\tiny{\mathcal{H}}}

\newtheorem{theorem}{Theorem}[section]
\newtheorem{corollary}[theorem]{Corollary}
\newtheorem{proposition}[theorem]{Proposition}
\newtheorem{remark}[theorem]{Remark}

\begin{document}
	
	\begin{frontmatter}
		
		\title{Integral Line-of-Sight Path Following Control of Magnetic Helical Microswimmers Subject to 
			Step-Out Frequencies} 
		
		\thanks[footnoteinfo]{This paper was not presented at any IFAC 
			meeting. Corresponding author A.~Mohammadi. Tel. +1-313-583-6787.}
		
		
		\author[Umich]{Alireza Mohammadi}\ead{amohmmad@umich.edu},    
		\author[UTD]{Mark W. Spong}\ead{mspong@utdallas.edu}               
		
		\address[Umich]{Department of Electrical \& Computer Engineering, University of Michigan, Dearborn}  
		\address[UTD]{Erik Jonsson School of Engineering \& Computer Science, University of Texas, Dallas}             


		\begin{abstract}                          
			This paper investigates the problem of straight-line path following for magnetic helical microswimmers. The control objective is to make the helical microswimmer to converge to a straight line without violating the step-out frequency constraint. The proposed feedback control solution is based on an optimal decision strategy (ODS) that is cast as a trust-region subproblem (TRS), i.e., a quadratic program over a sphere. The ODS-based control strategy minimizes the difference between the microrobot velocity and an integral  line-of-sight (ILOS)-based reference vector field while respecting the magnetic saturation constraints and ensuring the absolute continuity of the control input.  Due to the embedded integral action in the reference 
			vector field, the microswimmer will follow the desired straight line by compensating for the drift effect of the environmental disturbances as well as the microswimmer weight.  
		\end{abstract}
		
	\end{frontmatter}

\section{Introduction}
Swimming microrobots can be used  for  both \emph{in vivo} and \emph{in vitro} biomedical and 
micromanipulation applications. In \emph{in vivo} biomedical  applications, these robotic microswimmers can be employed for minimally invasive therapeutic and diagnostic procedures~\cite{ullrich2013mobility,cha2010electromagnetic}. In \emph{in vitro} or lab-on-a-chip  applications, these robots can be used for  protein-crystal handling~\cite{tung2013polymer} and cell manipulation/characterization~\cite{sakar2010single}.
%
Microrobots can be categorized based on their morphologies 
and actuators~\cite{abbott2009should}.  Among the two main classes 
of actuation methods  for microswimmers, i.e., 
\emph{untethered magnetic actuation}~\cite{abbott2009should,honda1996micro}  
and \emph{molecular motors}~\cite{behkam2006design}, using external 
magnetic fields for control of untethered 
microswimmers is more popular; mainly because the former scales well in terms of 
microfabrication and wireless power transmission/control.

The type of the microswimmer morphology is another 
factor that should be taken into account for microrobot design. Bead-like~\cite{yesin2006modeling}, 
eukaryotic-like~\cite{behkam2006design,khalil2016sperm}, and 
helical~\cite{bell2007flagella,mahoney2011velocity} shapes are the most widely-used morphologies for 
magnetic microswimmers. Magnetic microbeads are tiny rigid objects that are pulled through fluids using 
magnetic field gradients~\cite{abbott2009should}. In contrast to employing 
helical propellers or elastic tails, using magnetic gradient pulling is  far less efficient in terms of propulsion efficiency 
due to limitations on magnetic field sources. As shown by~\cite{abbott2009should}, there 
exists a microrobot size below which employing helical propellers and elastic tails is more efficient than pulling microbeads with field gradients. \\
Robotic microswimmers with eukaryotic-like morphologies move in their fluid environments by 
oscillating their flexible elastic tails.  On the other hand, in  
helical microswimmers, propulsive forces on the nanocoil structure are 
generated due to the rotation of the microrobot about the axis of its helix. Although the main 
mechanism for both elastic-tail and helical-propeller  magnetic microswimmers is based on 
transduction of magnetic torque to mechanical power, the helical morphology, 
which is inspired by bacterial flagella, has been shown to provide 
the best overall choice for \emph{in vivo} 
applications~\cite{abbott2009should}. Ease of direction reversal, 
independence of the microswimmer functionalization (e.g., drug coating) from fluid-dynamic properties, 
smooth transition from lumen to open environments, and the possibility of  
using non-uniform magnetic fields are among the principal reasons for superiority of 
helical propellers with respect to their elastic tail counterparts.  \\
%
Along with recent advances in microfabrication and actuation 
technologies, systematic design of automatic 
motion control algorithms for magnetic microswimmers 
has also been an active area of research~\cite{mahoney2011velocity,
	marino2013robust,arcese2013adaptive,fruchard2013estimation,mohammadi2018path,pedram2019optimal}. Being susceptible 
to  gravity and body fluid drag force 
uncertainties, limited accuracy and low localization rates in 
small scales, and limitations on electromagnetic actuators  are among the most significant 
challenges that arise in closed-loop control of magnetic microswimmers. The typical control approach in microrobot motion control 
literature relies on asymptotic tracking of suitably 
designed reference signals either in the form of a sequence of waypoints 
or in the form of time-based trajectories. Dahroug \emph{et al.}~\cite{dahroug2018some} have 
recently shown via experiments that  trajectory tracking control strategies are not robust to time delays and may create geometric deviations 
from the desired paths in low Reynolds swimming.  Indeed, as demonstrated 
by Aguiar \emph{et al.}~\cite{aguiar2005path}, 
trajectory tracking controllers for even LTI systems are subject to performance limitations in the presence of 
structural system constraints such as nonminimum phase zeros.  

Few researchers have proposed path following controllers for 
magnetic microswimmers~\cite{xu2015planar,oulmas2016closed,wu20193,oulmas20183d}. \textcolor{black}{The core of the underlying idea in~\cite{xu2015planar,oulmas2016closed,wu20193,oulmas20183d} relies on proper state transformations using the geometry of the target curved path in order to bring the dynamics of the microrobots into a chained form. Then, nonholonomic control techniques inspired by Samson and collaborators (see, e.g., \cite{samson1995control}) are invoked in order 
	to make the microrobot to converge to the target path with a desired velocity profile. In~\cite{oulmas2016closed}, the authors generalized the approach in~\cite{xu2015planar} from planar path following to 3D path following by devising nonholonomic control laws for higher dimension chained dynamics. In~\cite{oulmas20183d}, the Serret–Frenet frame considering the weight of the robot and lateral disturbances using
	the compensation inclination and direction angles has been employed. In~\cite{wu20193}, orientation-compensation model of the microrobot dynamics in the global coordinate frame
	is learned by proper backpropagation algorithms.} 
However, the proposed path following schemes do not directly address the control input saturation and the presence of unknown disturbances for the full dynamical model of the microswimmers.  \\
Magnetic microbead control is perhaps the most researched topic in the area of magnetic 
microswimmer control where a plethora of  control schemes  ranging from simple PID to adaptive backstepping 
controllers have been proposed in the literature. Marino \emph{et al.}~\cite{marino2013robust}
have employed robust $\mathcal{H}_\infty$ synthesis techniques to address steering of 
magnetic microbeads in motionless 
fluids under drag-force uncertainties and low image acquisition rates.  Fruchard \emph{et al.}~\cite{arcese2013adaptive} propose  
using an adaptive backstepping control law along with a high gain observer for  estimating the 
microbead velocity (see, also,~\cite{arcese2013adaptive}).

Unlike magnetic microbeads, very few control solutions  for eukaryotic-like microswimmers have been proposed in the literature (see, 
e.g.~\cite{kosa2007propulsion,khalil2016sperm}). Many of these solutions  rely on applying 
open-loop sinusoidal control inputs or designing linear controllers for Galerkin projection of the underlying 
elastic tail dynamical model. A recent promising approach to synthesis of propulsive gaits
for microswimmers with elastic tails is based on small-compliance assumptions and applying asymptotic 
perturbation techniques to the equations of motion~\cite{cicconofri2016motion}.

Early control solutions for magnetic helical microrobots have relied on considering the one-dimensional 
motion of the microswimmer along its helical axis~\cite{behkam2005modeling,bell2007flagella,fountain2010wireless,xu2015planar} 
without considering the effect of the microswimmer weight and/or other types of environmental 
disturbances. In a recent work, Mahoney \emph{et al.}~\cite{mahoney2011velocity} use 
an open-loop gravity compensation 
method for velocity control of helical magnetic microrobots that sink due to their 
own weights. The control-oriented model developed by  Mahoney \emph{et al.} does not assume neutral 
buoyancy of artificial helical 
microswimmers while directly taking into account the sinking effect of the microswimmer weight.  

In addition to the presence of disturbances, another major control challenge for magnetic 
helical microrobots is due to the existence of an upper limit on 
the robot rotational frequency around its helical axis. This threshold frequency, which 
is the maximum rotational frequency that keeps the robot in synchrony with the external 
rotating field, is known as the \emph{step-out frequency} beyond which the 
velocity of the microswimmer rapidly declines~\cite{abbott2009should,mahoney2014behavior}. In addition to 
rotating in sync with the magnetic field, most researchers also assume alignment
of the  field rotation axis with the microswimmer helical axis (see, e.g.,~\cite{mahoney2011velocity}). 

In this paper, we present a path following control law that formally 
guarantees practical convergence of magnetic microswimmers to desired straight lines 
with absolutely continuous velocity profiles while respecting the 
control input saturation limits in the presence of disturbances.  Our path following scheme uses an 
optimal decision strategy (ODS)-based control synthesis approach. ODS-based 
strategies belong to the larger family of optimization-based 
nonlinear controllers~\cite{morris2015continuity,
	Ames17a,bouyar2017}, whose applications in robotics and driverless 
cars are growing, thanks in part to recent advancements in mobile computation 
power.  Optimal decision strategies, which  were originally proposed in the context of 
controlling electric power systems and industrial robotic manipulators with bounded 
input~\cite{thomas1976model,spong1984control,spong1986control}, are \emph{pointwise optimal control laws} 
that minimize the deviation between the open-loop dynamics  vector field 
and a reference model vector field.  We propose using 
an integral line-of-sight 
(ILOS)-based reference vector field for our ODS-based control scheme inspired from the ILOS path following laws that are 
widely used for underactuated marine craft control~\cite{fossen2014line,caharija2016integral}). Our proposed ILOS-based guidance law incorporates the integral of the cross-track error of the 
microswimmer to the straight line.  

Since our proposed ODS-based QP, which computes control actions using the ILOS-based guidance law, has constraints on the 
magnitude of the control input vector, it belongs to the family of trust-region subproblems (TRS), i.e., QPs over  spheres and ellipsoids~\cite{adachi2017solving,hager2001minimizing}. TRS has long been of interest to the optimization research community (see, e.g., the classical work by Forsythe and Golub~\cite{forsythe1965stationary}); because a TRS needs to be solved in each step of trust-region optimization algorithms. 
In this paper, we provide the necessary and sufficient conditions for the existence of solutions to the special TRS that arises 
in the context of the magnetic microswimmer control problem. Furthermore, we provide sufficient conditions under which the absolute continuity of the generated control input is guaranteed. The absolute continuity of the control input is not only appealing from an existence and uniqueness of solutions perspective, but is also significant from a practical point of view. Indeed, it has been observed in practice that as long as a helical microswimmer is commanded a smooth desired-velocity profile with limits on acceleration, the \emph{highly correlated} requirements of rotating below the step-out frequency as well as alignment of the field rotation axis with 
the microrobot axis will be met~\cite{mahoney2011velocity}.

\noindent\textbf{Contributions of the paper.} This paper contributes to solving 
the path following control problem for swimming helical microbots in 
\emph{several ways}. First, the paper develops an ILOS-based guidance 
law for swimming microrobots, which is inspired from the automatic ship steering 
literature~\cite{caharija2012relative,fossen2014line,caharija2016integral}. In the 
presence of disturbances that drive the
microswimmer away from its desired path, embedding the integral 
compensation dynamics will build up a corrective action in the reference vector field. Second, 
using the ILOS-based guidance law, this paper casts the control input 
computation as a trust-region subproblem (TRS), i.e., a quadratic program over a sphere, 
which belongs to the wider class of real-time optimization-based controllers.  
In our previous work~\cite{mohammadi2018path}, we also used the ODS framework for designing 
path following controllers for  magnetic helical microswimmers.  While the proposed 
ODS-based controller in~\cite{mohammadi2018path}, which is based on the traditional 
line-of-sight (LOS) guidance law, respects the step-out frequencies, it relies on the full 
knowledge of the microswimmer dynamical parameters and assumes absence of disturbances. Additionally, 
the controller in~\cite{mohammadi2018path} formally guarantees neither the continuity of the control inputs nor the 
continuity of the commanded desired-velocity profiles. In this paper, we relax many of our 
prior assumptions.

The rest of this paper is organized as follows. First, we present the dynamical model of 
swimming helical microrobots in Section~\ref{sec:dynmodel}. Next, we formulate the 
straight-line path following control problem for a single swimming microrobot subject to control 
input constraints and outline our solution strategy in Section~\ref{sec:ContProb}. Thereafter, we present 
our ODS-based control scheme for swimming helical microrobots in Section~\ref{sec:Sol}. After 
presenting the simulation results in Section~\ref{sec:sims}, we 
conclude the paper with final remarks and 
future research directions in Section~\ref{sec:conc}.   

\noindent{\textbf{Notation.}} We let $\mathbb{R}_{+}$ denote the set of all non-negative real numbers. Given a vector 
$\mathbf{v}\in\mathbb{R}^3$ and two coordinate frames $\mathcal{A}$ and 
$\mathcal{B}$, we let  ${}^{\tiny\mathcal{A}} \mathbf{v}$ and ${}^{\tiny\mathcal{B}} \mathbf{v}$ be the coordinates 
of $\mathbf{v}$ in  $\mathcal{A}$ and  $\mathcal{B}$, respectively. Therefore, ${}^{\tiny\mathcal{A}} \mathbf{v} = {}^{\tiny\mathcal{A}}\mathbf{R}_{\tiny\mathcal{B}} {}^{\tiny\mathcal{B}} \mathbf{v}$, where ${}^{\tiny\mathcal{A}}\mathbf{R}_{\tiny\mathcal{B}}\in \text{SO}(3)$ is the rotation matrix from the 
frame $\mathcal{B}$ to the frame $\mathcal{A}$.   Given $\mathbf{x}\in\mathbb{R}^N$, we 
let $|\mathbf{x}|:=\sqrt{\mathbf{x}^{\top}\mathbf{x}}$ 
denote the Euclidean norm of $\mathbf{x}$. We let 
$\mathcal{B}_{r_0}(\mathbf{x}_0):=\big\{\mathbf{x} \in \mathbb{R}^N : |\mathbf{x}| < r_0 \big\}$ denote the ball centered at $\mathbf{x}_0$ with radius $r_0$. Given two vectors $\mathbf{v}_1$  and $\mathbf{v}_2$ and a coordinate 
frame $\mathcal{A}$, the dot product of the two vectors  is denoted by $\mathbf{v}_1 \cdot \mathbf{v}_2 := |\mathbf{v}_1| |\mathbf{v}_2| \cos(\theta)$, where 
$\theta$ is the angle between the two vectors $\mathbf{v}_1$  and $\mathbf{v}_2$. Hence, $\mathbf{v}_1 \cdot \mathbf{v}_2= {}^{\tiny\mathcal{A}}\mathbf{v}_1^\top {}^{\tiny\mathcal{A}}\mathbf{v}_2$. Given a square symmetric matrix $\mathbf{A}$, we let $\lambda_\text{min}(\mathbf{A})$ and $\lambda_\text{max}(\mathbf{A})$ denote 
the minimum and maximum eigenvalues of $\mathbf{A}$, respectively. Furthermore, we let 
$\mathbf{A}\succeq 0$ denote positive semi-definiteness of $\mathbf{A}$. We let $\mathbbm{1}:\mathbb{R}\to \{0,1\}$ denote the Heaviside step function, where $\mathbbm{1}(x)=0$ if $x\geq 0$, and $\mathbbm{1}(x)=0$ if $x<0$. 
Given a piecewise continuous function $\mathbf{u}:\mathbb{R}_{+} \to \mathbb{R}^m$, we 
let $\|\mathbf{u}\|_{\infty}:=\sup\limits_{t\in \mathbb{R}^+}\max\limits_{1\leq i \leq m} |u_i(t)|$.  

\section{Dynamical Model of Magnetic Helical Microswimmers}\label{sec:dynmodel} 
In this section we present the dynamics of magnetic helical microswimmers and summarize the 
results in~\cite{mahoney2011velocity}. 

\begin{figure}[t]
	\centering
	\begin{subfigure}{0.20\textwidth}
		\includegraphics[scale=0.20]{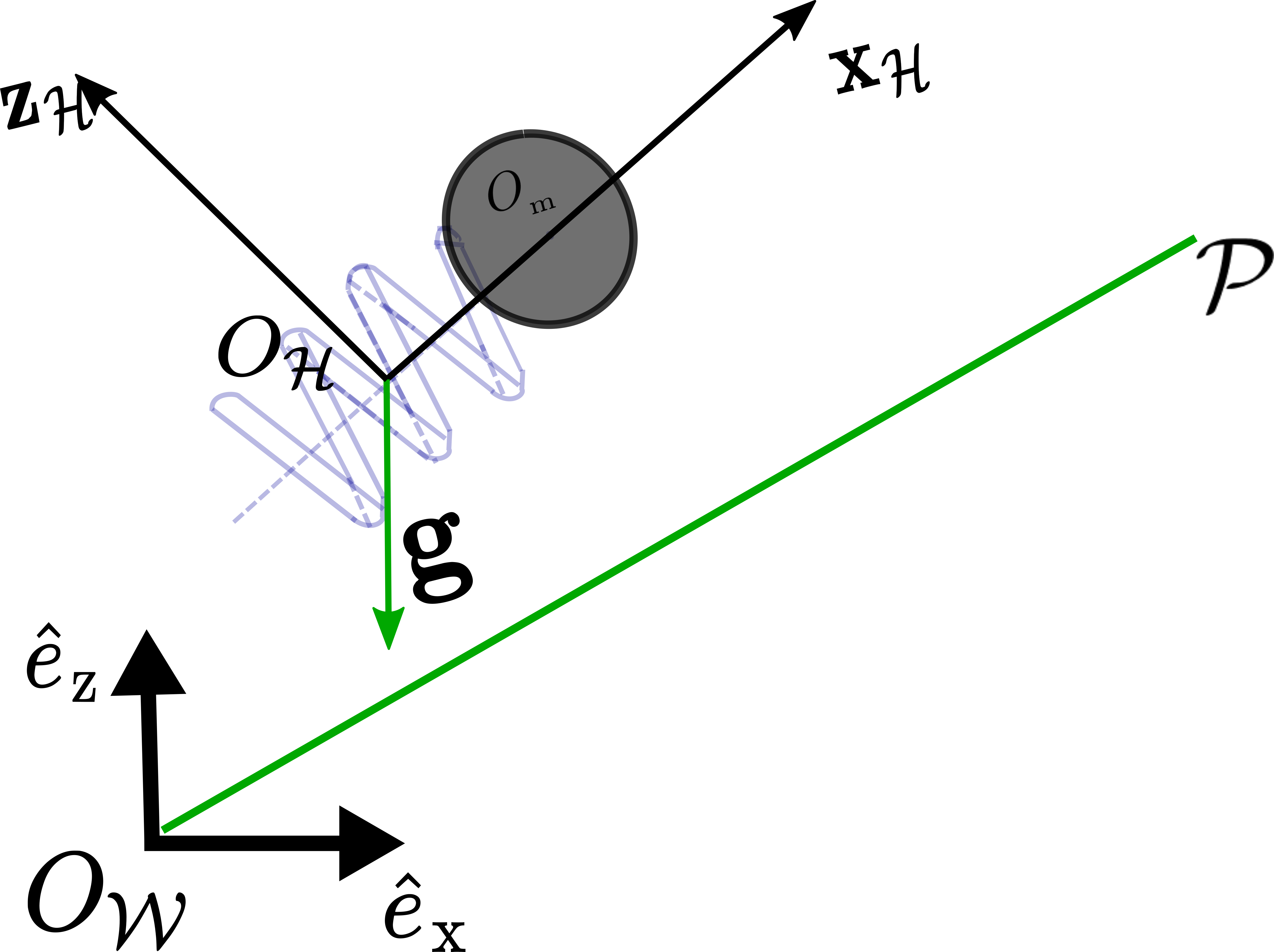}
		\caption{}
	\end{subfigure}
	\hspace{0.05in}
	\begin{subfigure}{0.23\textwidth}
		\includegraphics[scale=0.20]{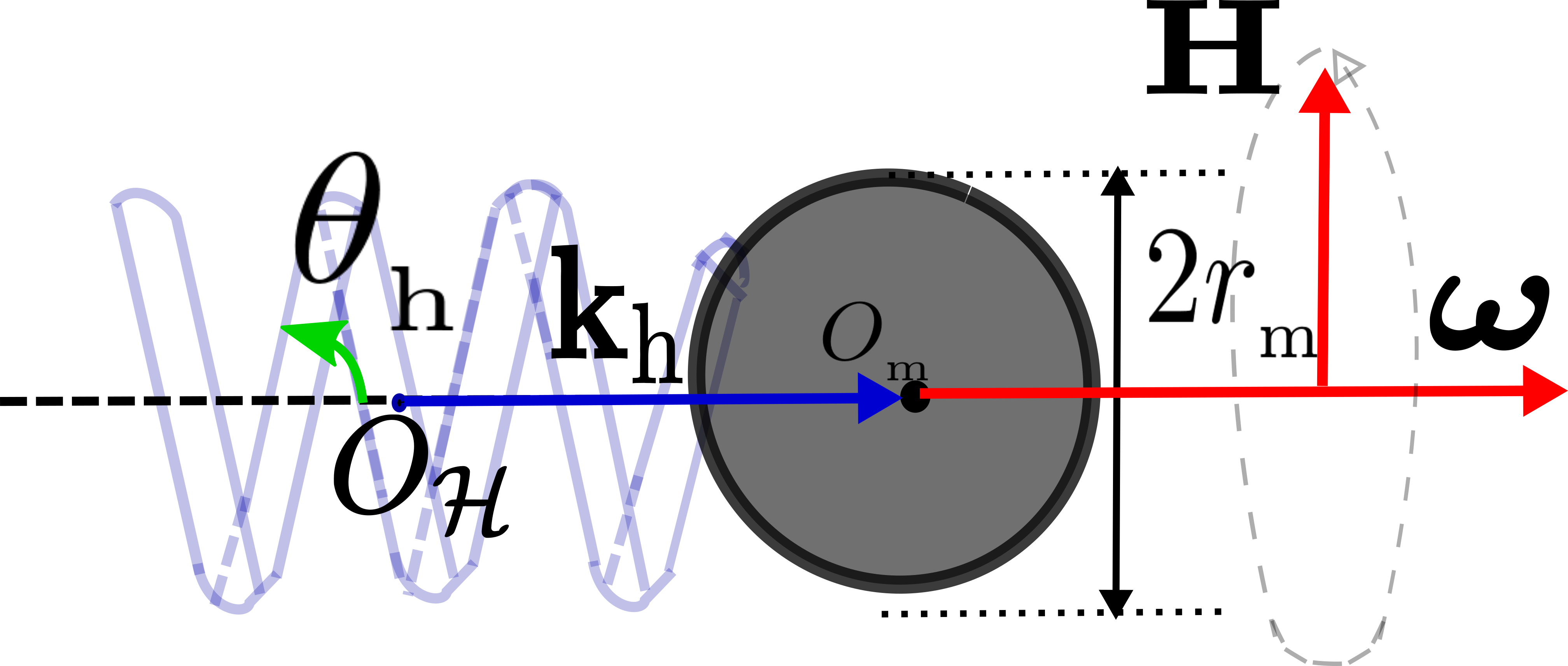}
		\caption{}
	\end{subfigure}
	\hspace{0.05in}
	\begin{subfigure}{0.20\textwidth}
		\includegraphics[scale=0.43]{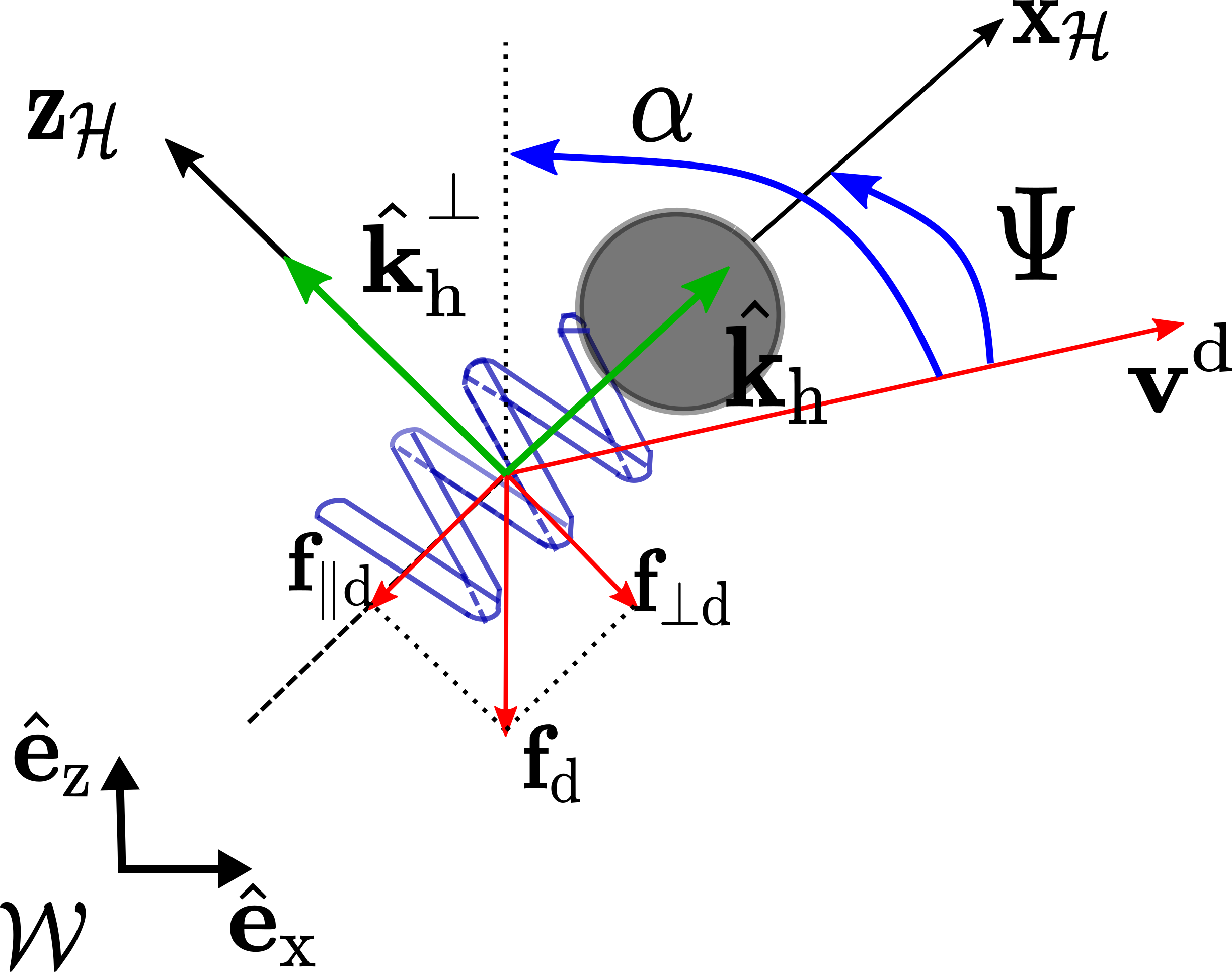}
		\caption{}
	\end{subfigure}
	\caption{The configuration of a magnetic helical microrobot.}
	\label{fig:micro}
\end{figure}

We consider the line $\mathcal{P}$ to which we would like the microswimmer to converge. We designate an arbitrary 
reference point $O_\mathcal{W}$ on $\mathcal{P}$ as the origin. Assuming that $\mathcal{P}$ is parallel to 
the direction vector $\hat{\vx{e}}_{{\tx{r}}}\in\mathbb{R}^3$, we have 
%
%
\begin{equation}\label{eq:line}
\mathcal{P}=\{ \vx{p}\in\mathbb{R}^3 : \vx{p} = \tau\hat{\vx{e}}_{\tx{r}} ,\, \tau\in \mathbb{R} \}. 
\end{equation}
Considering the gravitational acceleration 
vector $\vx{g}$, we define $\hat{\mathbf{e}}_\tx{x}$ to be the unit vector 
that is perpendicular to $\vx{g}$ and is contained in the 
plane spanned by $\mathcal{P}$ and $\vx{g}$. We fix the right-handed inertial coordinate  
frame $\mathcal{W}$ at $O_\mathcal{W}$ using the unit vectors $\hat{\mathbf{e}}_\tx{x}$, 
$\hat{\mathbf{e}}_\tx{z} := \tfrac{-\vx{g}}{|\vx{g}|}$, and 
$\hat{\mathbf{e}}_\tx{y} := \hat{\mathbf{e}}_\tx{x} \times \hat{\mathbf{e}}_\tx{z}$ (see Figure~\ref{fig:micro}(a)). 

\begin{remark}\label{rem:parallelPath}
	If the desired path $\mathcal{P}$ is parallel to $\vx{g}$, we choose another arbitrary line 
	$\mathcal{P}^\prime$  and fix the right-handed inertial coordinate frame $\mathcal{W}$ using 
	$\vx{g}$ and $\mathcal{P}^\prime$. 
\end{remark}
In addition to the inertial coordinate frame $\mathcal{W}$, we also consider the 
coordinate frame $\mathcal{H}$, whose origin $O\hfs$  is located at the helix center. We let the 
x-axis of $\mathcal{H}$, which we denote by $\vx{x}\hfs$, be aligned with the microrobot helical axis. 
We let the z-axis of the frame $\mathcal{H}$ be the axis $\vx{z}\hfs$ that is perpendicular to $\vx{x}\hfs$ 
and  is contained in the plane spanned 
by $\vx{x}\hfs$ and $\vx{g}$ (see Figure~\ref{fig:micro}(a)). The helical microswimmer geometry is completely determined by the number of turns of 
the helix $n_\tx{h}$, the helix pitch angle $\theta_\tx{h}$, the helix coil thickness $r_\tx{c}$, the helix radius $r_\tx{h}$, and the 
magnetic head radius $r_\tx{m}$ (see Figure~\ref{fig:micro}(b)).  We let $\mathbf{k}_\tx{h}$ 
denote the vector connecting the center of the helix  $O\hfs$ to the center of 
the magnetic head $O_\tx{m}$. Therefore, $\hat{\mathbf{k}}_\tx{h}:=\frac{\mathbf{k}_\tx{h}}{|\mathbf{k}_\tx{h}|}$ 
is parallel to the $\vx{x}\hfs$ axis. \textcolor{black}{In Figure~\ref{fig:micro}(b), the magnetic field 
	vector, which is induced by an external magnetic source, is shown by 
	vector $\mathbf{H}$. The rotation of the field vector $\mathbf{H}$ causes the magnetic helical microrobot 
	to rotate about its axis with an angular velocity vector given by 
	$\boldsymbol{\omega}$ shown in Figure~\ref{fig:micro}(b).} Figures~\ref{fig:micro}(a) and \ref{fig:micro}(b) depict the configuration of a 
generic helical microrobot as well as the coordinate frames $\mathcal{W}$ and $\mathcal{H}$.   
In this paper, we consider the wireless magnetic swimming of microrobots with helical propellers in low-Reynolds-number regimes. The Reynolds (Re) number,  which is used for studying propulsion mechanisms in fluidic environments, represents the ratio between the inertial forces and the resistive forces in a fluid. We denote the parallel and normal viscous drag force coefficients of the helical microswimmer by $\xi_{\perp}$ and $\xi_{\parallel}$, respectively. Furthermore, we let $\xi_{\text{vm}}$ denote the drag force coefficient of the microswimmer's spherical magnetic head.

The microrobot propulsion mechanism, which is based on transducing 
the external magnetic field energy to microrobot forward motion, can be described 
as follows. An external rotating uniform magnetic field, denoted by $\mathbf{H}$ in 
Figure~\ref{fig:micro}(b), causes the magnetic helical microrobot 
to rotate about its axis with an angular velocity vector given by 
$\boldsymbol{\omega}$. The resulting rotation about the helix axis in 
the ambient fluidic environment will then induce a screw-like motion and 
drive the microswimmer forward.  

In this paper, we use the control-oriented model developed in~\cite{mahoney2011velocity}. 
Using resistive force theory (RFT), Mahoney \emph{et al.}~\cite{mahoney2011velocity}  have 
derived the dynamical model of 3D helical microswimmers 
operating in low-Reynolds-number regimes. In this approach, the velocity of each 
infinitesimally small helix segment is mapped to parallel and perpendicular differential fluid drag forces 
acting on the segment. Integrating the differential forces along the length of the helix, the fluidic force and  
torque acting on the helical part of the robot are obtained. 
Adding the fluidic forces acting on the head, the 
dynamical equations of motion are obtained. In deriving their control-oriented dynamical model and later experimental implementations, Mahoney \emph{et al.}~\cite{mahoney2011velocity}  assume the following. 

\noindent{\textbf{H1)}}  The helical magnetic microswimmer rotates in synchrony with the external magnetic field. This 
assumption implies that the commanded rotation speed $|\boldsymbol{\omega}|$ is not above the step-out frequency $\Omega_{SO}$
of the microswimmer, beyond which the velocity of the microswimmer rapidly declines~\cite{mahoney2014behavior}. 

\noindent{\textbf{H2)}}  The central axis of the microswimmer $\mathbf{x}\hfs$  is always aligned with 
the magnetic field rotation axis. In other words, the angular velocity of the microswimmer can be directly 
commanded to be 
\begin{equation}\label{eq:contInput}
\boldsymbol{\omega} = \Omega \hat{\mathbf{k}}_{\text{h}}, 
\end{equation}
where $\Omega$ is the rotational frequency of the helical microswimmer about its axis. 

\begin{remark}\label{rem:contReqH1H2}
	It is known from the experimental observations (see, e.g.,~\cite{mahoney2011velocity,abbott2009should}) that as long as the microrobot is commanded a smooth velocity profile with limits on acceleration, the \emph{highly 
		correlated} requirements of rotating below the step-out frequency (required by \textbf{H1}) as well as alignment of 
	the field rotation axis with 
	the microrobot axis (required by \textbf{H2}) will be met in practice. As it will be shown in this paper, our control scheme formally guarantees 
	that the  commanded velocity respects both the step-out frequency constraint and the smoothness requirement. 
\end{remark}

In the inertial coordinate system $\mathcal{W}$, the velocity of the microswimmer $\vx{v}$ is related to 
applied non-fluidic forces $\vx{f}_\text{d}$  and the angular
velocity of 
the microswimmer $\boldsymbol{\omega}$, which represents the applied control input, through~\cite{mahoney2011velocity} 
\begin{equation}\label{eq:2D_dyn}
\wf\vx{v} = \wf \vx{D}\wf\vx{f}_\text{d} + \wf \vx{E} \wf\boldsymbol{\omega}, 
\end{equation}
where the matrices $\wf\vx{D}, \wf\vx{E}\in \mathbb{R}^{3\times 3}$  can be expressed using  
the following similarity transformations
\begin{equation}\label{eq:simTrans}
\wf\vx{D} = \wf\vx{R}\hfs \hf \vx{D} \hf\vx{R}\wfs,\; 
\wf \vx{E} = \wf\vx{R}\hfs \hf \vx{E} \hf\vx{R}\wfs.
\end{equation}
In~\eqref{eq:simTrans}, the matrix $\hf\vx{R}\wfs$ represents the rotational transformation matrix from
the inertial coordinate frame  $\mathcal{W}$ to the microswimmer coordinate frame $\mathcal{H}$. Furthermore, 
the two constant matrices $\hf\vx{D}, \hf\vx{E}$, which depend on the microswimmer's physical parameters are given by
\begin{equation}\label{eq:simTrans2}
\hf \vx{D} = \begin{bmatrix} \frac{1}{a_1} & 0 & 0 \\ 0 & \frac{1}{a_2} & 0 \\ 0 & 0 & \frac{1}{a_2} \end{bmatrix}, \; 
\hf \vx{E} = \begin{bmatrix} -\frac{b_{11}}{a_{1}} & 0 & -\frac{b_{13}}{a_{1}} \\ 0 & -\frac{b_{22}}{a_{2}} & -\frac{b_{23}}{a_{2}} \\ 0 & \frac{b_{23}}{a_{2}} & -\frac{b_{33}}{a_{2}} \end{bmatrix}. 
\end{equation}
The dependency of the constant parameters in~\eqref{eq:simTrans2} in terms of the 
parameters of the microswimmer, i.e., $\xi_{\parallel}$, $\xi_{\perp}$, 
$\theta_{\tx{h}}$, $r_\text{h}$, $n_\text{h}$, and $|\mathbf{k}_\text{h}|$, are summarized in Table~\ref{table:micro}. 

\begin{table*}[t]
	{
		\caption{The constant parameters in~\eqref{eq:simTrans2} in terms of physical characteristics of the magnetic helical microswimmer} 
		
		\centering 
		\makebox[\linewidth]{
			\begin{tabular}{|l | l| l| l|} 
				\hline  
				\textbf{Symbol} & \textbf{Description} & \textbf{Symbol} & \textbf{Description}\\ [0.1ex] 
				\hline 
				$a_{1}$ & $a_{\tx{h}1} + \xi_{\tx{vm}}$ & $a_{2}$ & $a_{\tx{h}2} + \xi_{\tx{vm}}$ \\
				$a_{\tx{h}1}$ & $\frac{2\pi n_\tx{h}  r_\tx{h} (\xi_{\parallel}\tx{c}^2_{\theta_\tx{h}}+\xi_{\perp}  \tx{s}^2_{\theta_\tx{h}})}{\tx{s}_{\theta_\tx{h}}}$ & $a_{\tx{h}2}$ & $\frac{\pi n_\tx{h} r_\tx{h} (\xi_{\perp}+\xi_{\parallel}\tx{s}^2_{\theta_\tx{h}}+\xi_{\perp} \tx{c}^2_{\theta_\tx{h}})}{\tx{s}_{\theta_\tx{h}}}$ \\
				$b_{11}$ & $2\pi n_\tx{h} r_\tx{h}^2 (\xi_{\parallel}-\xi_{\perp})\tx{c}_{\theta_\tx{h}}$ & 
				$b_{13}$ & $\frac{-b_{11}}{\tan(\theta_\tx{h})}$ \\
				$b_{22}$ & $\frac{-3b_{11}}{4}$ & 
				$b_{33}$ & $\frac{-b_{11}}{4}$  \\
				$b_{23}$ & $\xi_{\tx{vm}}| \mathbf{k}_\tx{h}|$ & & \\ 
				\hline 
			\end{tabular} 
		}
		\label{table:micro} 
	}
\end{table*} 
\begin{remark}\label{rem:sensingLim}
	As it can be seen from~\eqref{eq:2D_dyn}, one of the main challenges for closed-loop control of magnetic helical microswimmers is the need for sensing 
	the orientation of the robot about its central axis $\mathbf{x}_\mathcal{H}$. However, under \textbf{H1} and \textbf{H2}, it is possible to simplify~\eqref{eq:2D_dyn} in a way that there is no need for sensing the orientation of the microrobot. 
\end{remark}
The following proposition summarizes the main results in~\cite{mahoney2011velocity}. One of the major implications of this proposition 
is removing the need for sensing the orientation of the microrobot about its central axis. Furthermore, this proposition gives 
the direction and magnitude of the feedforward angular velocity vector command input that results in a desired velocity vector 
for the magnetic microswimmer\footnote{\textcolor{black}{The authors in~\cite{mahoney2011velocity} provide their computational arguments in a Section entitled ``Algorithm for Velocity Control with Gravity Compensation''. For the reader's convenience, we are succinctly presenting the computations and findings in~\cite{mahoney2011velocity} in the statement and proof of Proposition~\ref{prop:mahoney}.}} (see Figure~\ref{fig:micro}(c)). 


\begin{proposition}[\cite{mahoney2011velocity}]\label{prop:mahoney}
	Consider the magnetic microswimmer dynamics given by~\eqref{eq:2D_dyn}. Under Hypotheses \textbf{H1} and \textbf{H2}, 
	the dynamics of the helical microswimmer are equivalent to  
	\begin{equation}\label{eq:eq_dyn}
	\dot{\vx{p}} = e_{11}\vx{u} + {\vx{d}}_{\mu}, 
	\end{equation}
	where
	\begin{equation}\label{eq:notation}
	\dot{\vx{p}} := \wf\vx{v},\,\, \vx{u}:= \wf\boldsymbol{\omega},\,\,
	{\vx{d}}_{\mu} := \wf\vx{D}\wf\vx{f}_\text{d},
	\end{equation}
	and 
	\begin{equation}\label{eq:e11}
	e_{11}= \frac{-2\pi n_\tx{h} r_\tx{h}^2 (\xi_{\parallel}-\xi_{\perp})\tx{c}_{\theta_\tx{h}} \tx{s}_{\theta_\tx{h}} }
	{2\pi n_\tx{h}  r_\tx{h} (\xi_{\parallel}\tx{c}^2_{\theta_\tx{h}}+\xi_{\perp}  \tx{s}^2_{\theta_\tx{h}}) + \xi_{\tx{vm}}\tx{s}_{\theta_\tx{h}}}. 
	\end{equation}
	Furthermore, assume that the disturbance $\wf\vx{f}$ is contained in the plane spanned by $\mathcal{P}$ and 
	$\mathbf{g}$.  Given a desired velocity vector $\vx{v}^\text{d}$ in the plane spanned by $\mathcal{P}$ and 
	$\mathbf{g}$,  the feedforward angular velocity command input $\mathbf{u}$, which makes the angle
	\begin{equation}\label{eq:fdfwdDir}
	\Psi = \arctan\big(\tfrac{\tfrac{1}{a_2} |\vx{f}_\text{d}|\sin(\alpha) }{ |\vx{v}^\text{d}| + \tfrac{1}{a_2} |\vx{f}_\text{d}| \cos(\alpha)} \big), 
	\end{equation}
	with $\vx{v}^\text{d}$  and has the magnitude 
	\begin{equation}\label{eq:fdfwdMag}
	|\vx{u}| = \tfrac{|\vx{v}^\text{d}| \cos(\Psi) + \tfrac{1}{a_1} |\vx{f}_\text{d}|\cos(\alpha-\Psi)}{e_{11}}, 
	\end{equation}
	yields the closed-loop dynamics
	\begin{equation}\label{eq:closedLoop_dyn}
	\dot{\vx{p}}  = \vx{v}^\text{d}. 
	\end{equation}
\end{proposition}
%
%
\noindent{\textbf{Proof.}} Under Hypothesis \textbf{H2}, we have 
$\mathbf{u}=\Omega \hat{\mathbf{k}}_{\text{h}}$. Also, the coordinates of the vector $\hat{\mathbf{k}}_{\text{h}}$ in the microrobot frame $\mathcal{H}$ 
are given by $\hf\hat{\mathbf{k}}_{\text{h}}=[1,\,0,\,0]^{\top}$ (see Figure~\ref{fig:micro}(a)). Furthermore, from~\eqref{eq:simTrans2}, 
it can be seen that $\hf\mathbf{E}\hf\hat{\mathbf{k}}_{\text{h}}= e_{11} \hf\hat{\mathbf{k}}_{\text{h}}$, where $e_{11}=\tfrac{-b_{11}}{a_1}$. Since  
\[
\wf\mathbf{E}\wf\boldsymbol{\omega} = \Omega\wf\mathbf{E} \wf\hat{\mathbf{k}}_{\text{h}} = \Omega\wf\vx{R}\hfs \hf \vx{E} \hf\vx{R}\wfs \wf\hat{\mathbf{k}}_{\text{h}} =\] 
\[\Omega\wf\vx{R}\hfs \hf \vx{E} \hf \hat{\mathbf{k}}_{\text{h}} = \Omega e_{11} \wf\vx{R}\hfs \hf \hat{\mathbf{k}}_{\text{h}} = e_{11} (\Omega \wf\hat{\mathbf{k}}_{\text{h}}) = e_{11} \mathbf{u},
\]
the Equation given by~\eqref{eq:eq_dyn} holds. Considering~\eqref{eq:eq_dyn}, it can be seen that the feedforward control law
\begin{equation}\label{eq:tmp1}
\mathbf{u} = \tfrac{1}{e_{11}} \big\{ -\mathbf{d}_{\mu} + \vx{v}^\text{d} \big\}, 
\end{equation}
yields the closed-loop dynamics given by~\eqref{eq:closedLoop_dyn}. Since $\vx{f}_\text{d}$ 
is contained in $x_{\mathcal{H}}-z_{\mathcal{H}}$ plane (see Figure~\ref{fig:micro}(c)), we have
\begin{equation}\label{eq:fddecomp}
\vx{f}_\text{d} = (\vx{f}_\text{d}\cdot\hat{\vx{k}}_{\tx{h}}) \hat{\vx{k}}_{\tx{h}} + (\vx{f}_\text{d}\cdot
\hat{\vx{k}}^\perp_{\tx{h}})\hat{\vx{k}}^\perp_{\tx{h}}. 
\end{equation}
Moreover,  since the disturbance vector is contained in the plane spanned by $\mathcal{P}$ and $\vx{g}$, $\hf \vx{f}_\text{d}=\big[ \vx{f}_\text{d}\cdot\hat{\vx{k}}_{\tx{h}} ,\, \vx{f}_\text{d}\cdot
\hat{\vx{k}}^\perp_{\tx{h}} ,\, 0\big]^\top$ holds in the coordinate frame $\mathcal{H}$. Hence, from~\eqref{eq:simTrans2}, it can be seen that 
$\hf \vx{D} \hf \vx{f}_\text{d} = \big[ \tfrac{1}{a_1} \vx{f}_\text{d}\cdot\hat{\vx{k}}_{\tx{h}},\, \tfrac{1}{a_2} \vx{f}_\text{d}\cdot\hat{\vx{k}}_{\tx{h}}^\perp,\, 0\big]^\top$. Furthermore, we have 
%
$\mathbf{d}_{\mu}=\wf\vx{D}\wf\vx{f}_\text{d} = \wf\vx{R}\hfs \hf \vx{D} \hf\vx{R}\wfs \wf\vx{f}_\text{d} = \wf\vx{R}\hfs \hf \vx{D} \hf \vx{f}_\text{d}$. 
%
Therefore, the vector $\mathbf{d}_{\mu}$ can be written as 
\begin{equation}\label{eq:tmp3}
\mathbf{d}_{\mu}= \tfrac{1}{a_1}(\vx{f}_\text{d}\cdot\hat{\vx{k}}_{\tx{h}}) \hat{\vx{k}}_{\tx{h}} + \tfrac{1}{a_2}(\vx{f}_\text{d}\cdot 
\hat{\vx{k}}^\perp_{\tx{h}})\hat{\vx{k}}^\perp_{\tx{h}}. 
\end{equation}
Taking the dot product of $\mathbf{d}_{\mu}$ in~\eqref{eq:tmp3} with $\hat{\vx{k}}_{\tx{h}}$ and $\hat{\vx{k}}^\perp_{\tx{h}}$ yields $\mathbf{d}_{\mu}\cdot \hat{\vx{k}}_{\tx{h}}= \tfrac{1}{a_1} |\vx{f}_\text{d}| \cos( - \Psi + \alpha)$ and $\mathbf{d}_{\mu}\cdot \hat{\vx{k}}_{\tx{h}}^\perp= \tfrac{1}{a_2} |\vx{f}_\text{d}| \cos( \tfrac{\pi}{2} - \Psi + \alpha)$, respectively (see Figure~\ref{fig:micro}(c) for the angles $\Psi$ and $\alpha$). Also, $\vx{v}_\text{d} \cdot \hat{\vx{k}}_\text{h}=|\vx{v}^\text{d}| \cos(\Psi)$ and $\vx{v}_\text{d} \cdot \hat{\vx{k}}_\text{h}^\perp=|\vx{v}^\text{d}| \cos(\tfrac{\pi}{2}+\Psi)$. Since $\mathbf{u}= |\mathbf{u}| \hat{\vx{k}}_\text{h}$, taking the inner product of Equation~\eqref{eq:tmp1} with $\hat{\vx{k}}_\text{h}$ yields 
\begin{equation}\label{eq:tmp4}
|\mathbf{u}| = \tfrac{1}{e_{11}} \big\{ \tfrac{1}{a_1}|\vx{f}_\text{d}| \cos( - \Psi + \alpha) + |\vx{v}^\text{d}| \cos(\Psi) \big\}. 
\end{equation}
Therefore, Equation~\eqref{eq:fdfwdMag} holds. Similarly, taking the inner product of Equation~\eqref{eq:tmp1} with $\hat{\vx{k}}_\text{h}^{\perp}$ results in 
\begin{equation}\label{eq:tmp5}
0 = \tfrac{1}{e_{11}} \big\{ \tfrac{1}{a_2}|\vx{f}_\text{d}| \cos( \tfrac{\pi}{2} + \Psi - \alpha) + |\vx{v}^\text{d}| \cos(\tfrac{\pi}{2}+\Psi) \big\}. 
\end{equation}
Solving for $\Psi$ from~\eqref{eq:tmp4} and~\eqref{eq:tmp5} will result in~\eqref{eq:fdfwdDir}. \hfill$\blacksquare$

According to Proposition~\ref{prop:mahoney}, which holds under \textbf{H1} and \textbf{H2}, we can constrain the 
motion of the magnetic microswimmer to the plane containing $\mathcal{P}$ and $\vx{g}$ by commanding the 
angular velocity vector $\boldsymbol{\omega}$ to be in this plane. Therefore, throughout the rest of the paper, 
we consider the motion of the helical microswimmer to be in the $\hat{\mathbf{e}}_\text{x}-\hat{\mathbf{e}}_\text{z}$  plane. 
Consequently,  the position of the center of mass 
and the velocity of the microswimmer in  the $\hat{\mathbf{e}}_\tx{x}\text{-}\hat{\mathbf{e}}_\tx{z}$ plane
are given by $\mathbf{p}=[p_\text{x},\,p_\text{z}]^{\top}$ and $\vx{v}=\dot{\mathbf{p}}$, respectively, where $\mathbf{v}=[v_\text{x},\,v_\text{z}]^{\top}$. We denote the angle of 
the center of mass (COM) position vector in the $\hat{\mathbf{e}}_\tx{x}\text{-}\hat{\mathbf{e}}_\tx{z}$ plane by $\theta$. 
Therefore, 
\begin{equation}\label{eq:theta}
\theta := \text{atan2}(p_\text{z},p_\text{x}). 
\end{equation}
\begin{remark}\label{rem:}
	Under~\textbf{H2},  we can directly command $\boldsymbol{\omega}$, which is aligned 
	with the axis of the helical microrobot, and hence there is no need for measuring the orientation of the helical 
	microswimmer.  The only required measurements for implementing our proposed control laws are the COM 
	Cartesian coordinates given by  $p_\text{x}$ and $p_\text{z}$. The angle $\theta$ in~\eqref{eq:theta} 
	and the magnitude $|\mathbf{p}|$ are the polar coordinates of the 
	position vector $\mathbf{p}=[p_\text{x},\,p_\text{z}]^{\top}$. 
\end{remark}
%

\section{Control Problem Formulation and Solution Strategy}
\label{sec:ContProb}

In this section we formulate the straight-line 
path following control problem for swimming magnetic
helical microrobots and outline our solution 
strategy. 

Before stating the control objective, we introduce a coordinate transformation that maps 
the position of the microswimmer's COM to the cross-track error to the path and the 
projected position along the path, respectively. In particular, given the straight 
line $\mathcal{P}$ in~\eqref{eq:line} and assuming that it makes 
the angle $\theta_\text{r}$ with $\hat{\mathbf{e}}_\text{x}$ (see Figure~\ref{fig:micro}(a)), 
we define the cross-track error  
to $\mathcal{P}$ as 
\begin{equation}\label{eq:output}
\varepsilon := \hat{\mathbf{e}}_{_{\theta_\tx{r}}}^{\perp^\top} \mathbf{p},
\end{equation}
\noindent where 
\begin{equation}\label{eq:ehatThetarPerp}
\hat{\mathbf{e}}_{_{\theta_\tx{r}}}^{\perp} :=  \begin{bmatrix} -\sin(\theta_\tx{r}) \\  \cos(\theta_\tx{r}) \end{bmatrix},
\end{equation}
is the unit vector perpendicular to $\mathcal{P}$. We also define the projected position along the path 
as 
\begin{equation}\label{eq:zeroState}
z := \hat{\mathbf{e}}_{_{\theta_\tx{r}}}^{\top} \mathbf{p},
\end{equation}
\noindent where 
\begin{equation}\label{eq:ehatThetar}
\hat{\mathbf{e}}_{_{\theta_\tx{r}}}^\top :=  \begin{bmatrix} \cos(\theta_\tx{r}),\;  \sin(\theta_\tx{r}) \end{bmatrix},
\end{equation}
is the unit vector parallel to the line $\mathcal{P}$. When $\varepsilon=0$, the variable $z$ provides the position of 
the microswimmer along $\mathcal{P}$. From~\eqref{eq:theta}, it can be seen that $p_\text{x}=|\vx{p}| \cos(\theta)$ and $p_\text{z}=|\vx{p}| \sin(\theta)$. Hence, we can rewrite the cross-track error as 
$\varepsilon = \hat{\mathbf{e}}_{_{\theta_\tx{r}}}^{\perp^\top} \mathbf{p} = \big[ -\sin(\theta_\tx{r}) p_\text{x} +  \cos(\theta_\tx{r}) p_\text{z} \big] =$
$| \vx{p} | \big[ \sin(\theta)\cos(\theta_\tx{r}) - \cos(\theta)\sin(\theta_\tx{r}) \big]$.
Since $\sin(\theta-\theta_{\tx{r}})=\sin(\theta)\cos(\theta_\tx{r}) - \cos(\theta)\sin(\theta_\tx{r})$, we have
\begin{equation}\label{eq:output2}
\varepsilon = | \vx{p} | \sin(\Delta \theta),
\end{equation}
where $\Delta \theta := \theta - \theta_\tx{r}$. Similarly, it can be shown that 
\begin{equation}\label{eq:z2}
z = | \vx{p} | \cos(\Delta \theta).
\end{equation}
\textbf{Straight-Line Path Following Control (LFC) Problem.} Consider a given step-out frequency 
$\Omega_{\text{SO}}>0$ and the straight line $\mathcal{P}$ given by~\eqref{eq:line}. Consider the 
planar magnetic microswimmer whose dynamics are given by~\eqref{eq:eq_dyn}. Assume that the 
disturbance input $\vx{d}_\mu: \mathbb{R}_+ \to \mathbb{R}^2$ in~\eqref{eq:notation} is piecewise 
continuous and satisfies $\|\vx{d}_\mu\|_\infty< d^\ast$ for a positive, yet 
unknown, $d^\ast$.  Make the cross-track error $\varepsilon$ in~\eqref{eq:output2} to  
practically converge\footnote{Practical stabilization of a variable means that by a suitable choice of
	controller parameters the variable is made to converge to an arbitrarily small
	neighborhood of its desired value.} to $\mathcal{P}$  with a continuous and bounded velocity profile $\dot{\vx{p}}(t)$ such that $|\vx{u}(t)|\leq \Omega_{\text{SO}}$ for all $t\geq 0$. 
\begin{remark}\label{rem:distRem}
	In both the work of Mahoney \emph{et al.}~\cite{mahoney2011velocity} and our prior work 
	in~\cite{mohammadi2018path}, it is assumed that the only disturbance acting on the microswimmer 
	is the weight of the microrobot and its magnitude is known. In this paper, we remove this requirement. 
\end{remark}

\noindent\textbf{Solution Strategy.}  Our solution to microrobot LFC 
problem unfolds in the following steps.
\begin{itemize}
	\item[] \textbf{Step 1:} In Section~\ref{subsec:los}, we consider the reference model  
	\begin{equation}
	\label{eq:refmodel}
	\dot{\vx{p}} = \vx{v}^{\tx{d}}(\vx{p},s), 
	\end{equation}
	where $\vx{v}^{\tx{d}}(\vx{p},s)$ is the desired closed-loop 
	vector field coming from an ILOS guidance law, which depends on a 
	dynamic variable $s$. In Step~1, we prove that the LFC problem objective is achieved 
	when the velocity command input $\mathbf{u}$ is set equal to the reference 
	vector field in~\eqref{eq:refmodel} in the absence of disturbances and step-out frequency 
	constraints. Furthermore, we show that under 
	bounded disturbances the cross-track error and the dynamic variable $s$ in the ILOS guidance 
	law remain bounded. 
	\item[] \textbf{Step 2:}  Having obtained a desired closed loop vector field 
	that achieves the LFC objective in Step~1, we cast the control problem as an ODS-based 
	quadratic program over a sphere in Section~\ref{subsec:ods}. This ODS-based quadratic program 
	minimizes,  at each position $\vx{p}$, the difference between the open-loop and the reference 
	model vector fields in~\eqref{eq:refmodel}, while 
	respecting the step-out frequency constraints on the angular velocity command inputs. 
	In this step, we provide closed-form solutions for the ODS-based 
	QP and give sufficient conditions under which the generated angular velocity command input 
	is smooth. 
	%
	%
\end{itemize}
%

\section{Straight-Line Path Following Control Problem Solution}
\label{sec:Sol}

In order to solve LFC problem, we proceed according to the solution strategy 
outlined in the previous section. 

\subsection{Step 1: Integral line-of-sight reference vector field}
\label{subsec:los}
\begin{figure}[t]
	\begin{subfigure}{0.18\textwidth}
		\includegraphics[scale=0.30]{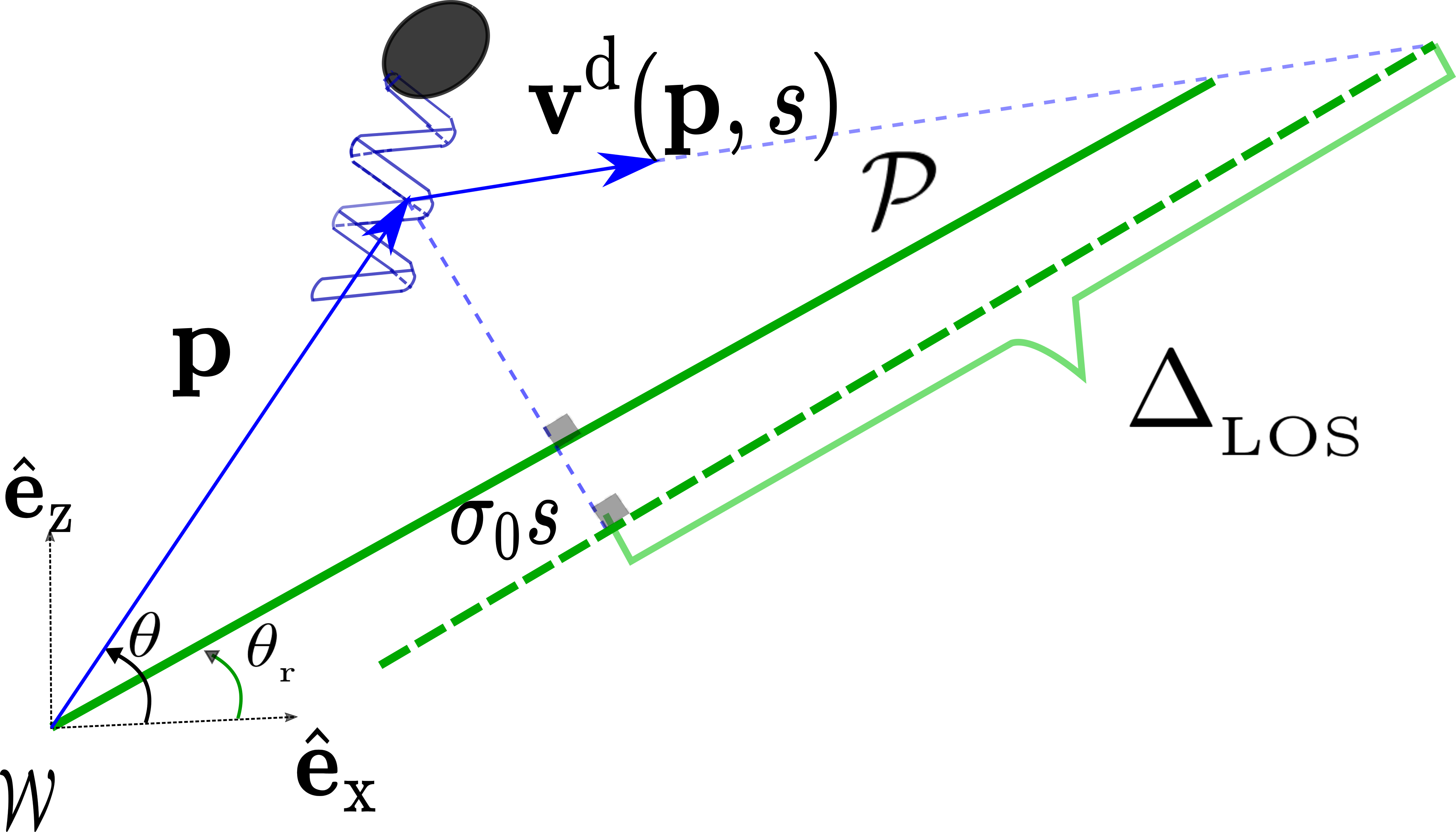}
		\caption{}
	\end{subfigure}
	\hspace{0.2in}
	\begin{subfigure}{0.15\textwidth}
		\includegraphics[scale=0.25]{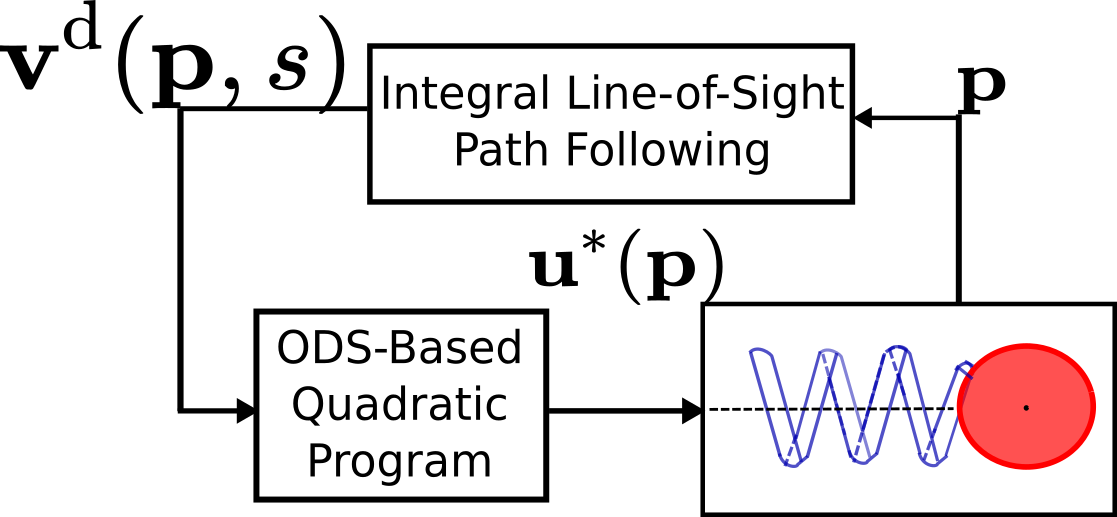}
		\vspace{0.11in}
		\caption{}
	\end{subfigure}
	\caption{(a) Integral line-of-sight reference vector field and (b) block diagram of the ODS-based control scheme.}
	\label{fig:los}
\end{figure}

Considering the path $\mathcal{P}$ in~\eqref{eq:line}, 
we propose a reference vector field for the closed-loop 
dynamics that achieves the LFC problem objective. This vector 
field is inspired from the ILOS path following laws for 
underactuated marine craft control 
(see, e.g.,~\cite{caharija2016integral}). In ILOS-based guidance schemes, the 
cross-track error of the  moving 
object to the desired path is minimized while the controlled 
object is pointing at a moving target point on the desired path 
(see Figure~\ref{fig:los}(a)). In addition to the cross track-error to $\mathcal{P}$, 
ILOS-based guidance laws incorporate the integral of the cross-track error using a 
dynamic variable $s$. In the presence of disturbances that drive the
microswimmer away from its desired path, embedding the integral 
compensation dynamics via the dynamic variable $s$ will build up a corrective action 
in the reference vector field. 
Given the LFC problem for the magnetic microswimmer, we propose the following 
ILOS reference vector field
\begin{equation}\label{eq:los}
\vx{v}^{\tx{d}}(\vx{p},s) = \alpha_\text{d}\vx{R}_{\theta_\tx{r}} \begin{bmatrix} \Delta_\tx{LOS} \\ -| \vx{p}| \sin(\Delta \theta) - 
\sigma_0 s \end{bmatrix},
\end{equation}
\noindent where $\alpha_\text{d}>0$ and the integral gain $\sigma_0>0$ are constant design parameters, $\vx{p}$ is the position of the center of mass of the microrobot with respect to the inertial frame $\mathcal{W}$,   and 
the matrix $\mathbf{R}_{\theta_{\tx{r}}}$ is the rotation matrix by
$\theta_{\tx{r}}$. The parameter $\Delta_\tx{LOS}>0$, which determines the point along 
$\mathcal{P}$ at which the microrobot should be pointed (see Figure~\ref{fig:los}(a)), is called 
the \emph{look-ahead distance}. Furthermore, $s$ is a dynamic variable whose dynamics are governed by 
\begin{equation}\label{eq:losdynvar}
\dot{s} = -k_\text{d}s + 
\Delta_{\tx{LOS}}\frac{| \vx{p}| \sin(\Delta \theta)}{(| \vx{p}| 
	\sin(\Delta \theta)+\sigma_0 s)^2 + \Delta_{\tx{LOS}}^2},
\end{equation}
where the damping gain $k_\text{d}>0$ is a constant 
design parameter. Figure~\ref{fig:los}(a) provides the 
geometric interpretation of the proposed ILOS guidance law in~\eqref{eq:los} and~\eqref{eq:losdynvar}.  
The idea behind~\eqref{eq:los} and~\eqref{eq:losdynvar} is that
the integral of the cross-track error $\varepsilon=|\mathbf{p}| \sin(\Delta \theta)$ will 
allow the angle between $\vx{v}^{\tx{d}}(\vx{p},s)$ and $\mathcal{P}$  to be
non-zero when $\varepsilon=0$, i.e., when the microswimmer is moving on the desired
line. In particular, in the presence of disturbances driving the
microswimmer away from its path, the integral of the cross-track error
will build up to create a non-zero side-slip angle to follow the line.

We can rewrite the ILOS reference vector field in~\eqref{eq:los} and~\eqref{eq:losdynvar} in terms of the cross-track error given 
by~\eqref{eq:output2} as 
\begin{equation}\label{eq:los2}
\begin{aligned}
\vx{v}^{\tx{d}}(\varepsilon,s) = \alpha_\text{d} \vx{R}_{\theta_\tx{r}} \begin{bmatrix} \Delta_\tx{LOS} \\ -\varepsilon - 
\sigma_0 s \end{bmatrix}, \\
\dot{s} = -k_\text{d}s + \Delta_{\tx{LOS}}\frac{\varepsilon}{(\varepsilon+\sigma_0 s)^2 + \Delta_{\tx{LOS}}^2}. 
\end{aligned}
\end{equation}

We now derive the closed-loop dynamics of the cross-track error under the effect of disturbances while assuming that 
there are no restrictions on the control input magnitude. In the next section, we formally take into account the angular velocity 
command input restrictions due to the step-out frequency constraint. 

Considering the microswimmer's  dynamics in \eqref{eq:eq_dyn}, an estimate $\hat{e}_{11}$ of the physical parameter 
$e_{11}$ in~\eqref{eq:e11} and an estimate $\hat{\mathbf{d}}_\mu$ of  the disturbance vector $\mathbf{d}_\mu$ in~\eqref{eq:notation}, 
and assuming that the control input has been chosen to be 
\begin{equation}\label{eq:nomCont}
\mathbf{u}= \frac{1}{\hat{e}_{11}}\vx{v}^\text{d} - \frac{1}{\hat{e}_{11}} \hat{\mathbf{d}}_\mu, 
\end{equation}
we get the closed-loop dynamics
\begin{equation}\label{eq:closed1}
\dot{\vx{p}} = \tfrac{e_{11}}{\hat{e}_{11}} \vx{v}^\text{d} + \vx{d}_\mu - \tfrac{e_{11}}{\hat{e}_{11}} \hat{\mathbf{d}}_\mu. 
\end{equation}

\begin{remark}\label{rem:Estimates}
	If no knowledge of $e_{11}$ and $\mathbf{d}_\mu$ are available, one can choose the 
	estimates $\hat{e}_{11}=1$ and $\hat{\mathbf{d}}_\mu=\mathbf{0}$ in~\eqref{eq:nomCont}. 
	In deriving the cross-track error dynamics under ILOS-based guidance laws and their 
	stability properties, we assume 
	that  $\hat{e}_{11}=1$ and $\hat{\mathbf{d}}_\mu=\mathbf{0}$. 
\end{remark}

Therefore, the dynamics of the cross-track error in~\eqref{eq:output2} under the control law in~\eqref{eq:nomCont}  are given by 
\begin{equation}\label{eq:varepsilonDot}
\dot{\varepsilon}= e_{11}\alpha_{\tx{d}}\hat{\vx{e}}_{_{\theta_\tx{r}}}^{\perp^\top} \dot{\vx{p}} = 
\alpha_{\tx{d}}\hat{\vx{e}}_{_{\theta_\tx{r}}}^{\perp^\top}\big(\tfrac{e_{11}}{\hat{e}_{11}} \vx{v}^\text{d} + \vx{d}_\mu - \tfrac{e_{11}}{\hat{e}_{11}} \hat{\mathbf{d}}_\mu\big). 
\end{equation}
Using the ILOS guidance law $\vx{v}^{\text{d}}=\vx{v}^{\tx{d}}(\varepsilon,s)$ in~\eqref{eq:los2}, assuming $\hat{e}_{11}=1$ and 
$\hat{\mathbf{d}}_\mu=\mathbf{0}$ (see Remark~\ref{rem:Estimates}), and noticing that $\hat{\vx{e}}_{_{\theta_\tx{r}}}^{\perp^\top}\vx{R}_{\theta_\tx{r}}=[0,\,1]$, we obtain
\begin{equation}\label{eq:zdyn}
\dot{\varepsilon}= \alpha_{\tx{d}}e_{11}(-\varepsilon - \sigma_0 s) + \alpha_\text{d} d_\mu^{\perp}, 
\end{equation}
\noindent where $d_\mu^{\perp}:=\hat{\vx{e}}_{_{\theta_\tx{r}}}^{\perp^\top} {\vx{d}}_{\mu}$.  
Similarly, the dynamics of $z$ defined in~\eqref{eq:zeroState} are found to be
\[
\dot{z}= e_{11}\alpha_{\tx{d}}\Delta_{\text{LOS}} + \alpha_\text{d} d_\mu^{\parallel}, 
\]
where $d_\mu^{\parallel}:=\hat{\vx{e}}_{_{\theta_\tx{r}}}^{\top} {\vx{d}}_{\mu}$. Consequently, the closed-loop dynamics 
of the helical microswimmer under the control law~\eqref{eq:nomCont} with $\hat{e}_{11}=1$ and 
$\hat{\mathbf{d}}_\mu=\mathbf{0}$ (see Remark~\ref{rem:Estimates}) and the ILOS guidance law $\vx{v}^{\text{d}}=\vx{v}^{\tx{d}}(\varepsilon,s)$ in~\eqref{eq:los2}  are governed by 
\begin{equation}\label{eq:los3}
\begin{aligned}
\dot{\varepsilon} = -\alpha_{\tx{d}} \varepsilon - \alpha_{\tx{d}}\sigma_0 s +\alpha_{\tx{d}} d_\mu^{\perp}, \\
\dot{s} = -k_\text{d}s + \Delta_{\tx{LOS}}\frac{\varepsilon}{(\varepsilon+\sigma_0 s)^2 + \Delta_{\tx{LOS}}^2},\\
\dot{z} =  \alpha_{\tx{d}}e_{11}\Delta_{\text{LOS}} + \alpha_\text{d} d_\mu^{\parallel}. 
\end{aligned}
\end{equation}

\begin{remark}\label{rem:LOSprop}
	As it can be seen from Equation~\eqref{eq:los3}, the dynamics of $s$ have 
	the property that $\dot{s} \to -k_\text{d}s$ as $\varepsilon \to \infty$, implying that 
	the rate of integration will decrease with large cross-track errors. In particular, the
	integral term will be less dominant when the cross-track error
	is large, i.e., when the microswimmer is far from the desired line $\mathcal{P}$. As stated 
	in~\cite{caharija2016integral}, such 
	error-dependent attenuation of integral action  will reduce the risk of 
	integrator wind-up and its resulting performance limitations.
\end{remark}

In order to investigate the stability properties of the cross-track error dynamics in~\eqref{eq:los3}, we define the following 
state and disturbance vectors

\begin{equation}\label{eq:xCross}
\vx{x}:= \begin{bmatrix} \varepsilon \\  s \end{bmatrix}, 
\, \vx{d}_\mu^{\perp}:= \begin{bmatrix} d_\mu^{\perp} \\  0 \end{bmatrix}. 
\end{equation}

Using~\eqref{eq:los3}, the dynamics of $\vx{x}$ can be written as 
\begin{equation}\label{eq:xCrossdyn}
\dot{\vx{x}} = \vx{A} \vx{x} + \vx{G}(\vx{x})\vx{x} + \alpha_\text{d} \vx{d}_\mu^{\perp}, 
\end{equation}
\noindent where 
\begin{equation}\label{eq:xCrossdyn2}
\begin{aligned}
\vx{A}  := -\alpha_\text{d} \begin{bmatrix}  1 & \sigma_0 \\ 0 & \frac{k_\text{d}}{\alpha_\text{d}} \end{bmatrix},&
\vx{G}(\vx{x})  := \frac{\Delta_{\tx{LOS}}}{\vx{x}^\top \vx{H} \vx{x} + \Delta_{\tx{LOS}}^2}\vx{B},\\ \textcolor{black}{\vx{H}  := \begin{bmatrix}  1 & \sigma_0  \\ \sigma_0 & \sigma_0^2 \end{bmatrix}}, &  
\vx{B}  := \begin{bmatrix}  0 & 0  \\ 1 & 0 \end{bmatrix}.
\end{aligned}
\end{equation}
The following proposition describes the stability properties of the cross-track error dynamics in~\eqref{eq:xCrossdyn} and~\eqref{eq:xCrossdyn2} as well as 
the steady-state velocity along $\mathcal{P}$. 
\begin{proposition}\label{prop:los}
	Consider the helical microswimmer dynamics in~\eqref{eq:eq_dyn} under the 
	angular velocity control input~\eqref{eq:nomCont} with  $\hat{e}_{11}=1$, $\hat{\mathbf{d}}_\mu=\mathbf{0}$, and the ILOS guidance law $\vx{v}^{\text{d}}=\vx{v}^{\text{d}}(\mathbf{p},s)$ 
	given by~\eqref{eq:los} and~\eqref{eq:losdynvar}. Suppose that there exist symmetric positive definite matrices $\mathbf{\Gamma}$ and
	$
	\mathbf{P}=\begin{bmatrix} p_{11} & p_{12} \\ p_{12} & p_{22} \end{bmatrix}\; 
	$ 
	such that 
	\begin{equation}\label{eq:stabCond}
	\begin{aligned}
	\vx{A}^\top \vx{P} + \vx{P} \vx{A} = -\vx{\Gamma},\\
	\lambda_{\text{min}}(\vx{\Gamma}) > \frac{p_{12} + \sqrt{p_{12}^2+p_{22}^2}}{\Delta_{\text{LOS}}}.  
	\end{aligned}   
	\end{equation}
	Then, under 
	$d_\mu^{\perp}=0$, the cross-track error dynamics 
	in~\eqref{eq:xCrossdyn} are globally exponentially stable (GES). Moreover, if $d_\mu^{\parallel}=0$, $\dot{\vx{p}}(t) \to 
	e_{11}\alpha_{\tx{d}}\Delta_{\tx{LOS}}\hat{\mathbf{e}}_{_{\theta_\text{r}}}$ as $t\to \infty$. Furthermore, if 
	$\lambda_{\text{min}}(\vx{\Gamma}) > \tfrac{p_{12} }{\Delta_{\text{LOS}}} + \sqrt{p_{12}^2+p_{22}^2}(\alpha_\text{d} + \tfrac{1}{\Delta_{\text{LOS}}})$, the cross-track error dynamics in~\eqref{eq:xCrossdyn} are input-to-state stable (ISS). Consequently, if  $\|\vx{d}_\mu\|_\infty < d^\ast$ for some constant $d^\ast>0$, then 
	$[\varepsilon, \, s]^\top$ will converge to the ball $\mathcal{B}_{ d^\ast\lambda_{\text{max}}(\mathbf{P})/\lambda_{\text{min}}(\mathbf{P})}(\mathbf{0})$.  
\end{proposition}

\textbf{Proof.}  Consider the quadratic Lyapunov function candidate 
\begin{equation}\label{eq:Vcandid}
V(\vx{x}) := \vx{x}^\top \vx{P} \vx{x},
\end{equation}
and compute its derivative $\dot{V}(\vx{x}) :=  \tfrac{\partial V}{\partial \vx{x}}\dot{\vx{x}}$ 
along the trajectories of the error dynamics in~\eqref{eq:xCrossdyn} to obtain 
\begin{equation}\label{eq:Vdot}
\begin{aligned}
\dot{V}(\vx{x})& = \vx{x}^\top (\vx{A}^\top \vx{P} + \vx{P} \vx{A}) \vx{x} + 
\frac{\Delta_{\tx{LOS}}}{\vx{x}^\top \vx{H} \vx{x} + \Delta_{\tx{LOS}}^2} \vx{x}^\top (\vx{B}^\top \vx{P} + \\
& \vx{P} \vx{B}) \vx{x}  + 2\alpha_\text{d} \vx{x}^\top \vx{P} \vx{d}_{\mu}^{\perp}.
\end{aligned}
\end{equation}
Using the Rayleigh's inequality, we have 
$\label{eq:Vdotineq}
\dot{V}(\vx{x}) \leq -\lambda_{\text{min}}(\vx{\Gamma}) |\vx{x}|^2 
+  \frac{\Delta_{\tx{LOS}} \lambda_{\text{max}}(\vx{B}^\top \vx{P} + \vx{P} \vx{B})}{\lambda_{\text{min}}(\vx{H}) |\vx{x}|^2  + \Delta_{\tx{LOS}}^2} |\vx{x}|^2 
+ 2 \alpha_\text{d} \vx{x}^\top \vx{P} \vx{d}_{\mu}^{\perp}$.
Since \textcolor{black}{$\lambda_{\text{min}}(\mathbf{H})=0$}, $\lambda_{\text{max}}(\vx{B}^\top \vx{P} + \vx{P} \vx{B})=p_{12} + \sqrt{p_{12}^2+p_{22}^2}$, and from~\eqref{eq:stabCond}, it 
can be seen that 
\begin{equation}\label{eq:Vdotineq2}
\dot{V}(x) \leq -\big(\lambda_{\text{min}}(\vx{\Gamma}) - \tfrac{p_{12} + \sqrt{p_{12}^2+p_{22}^2}}{\Delta_{\tx{LOS}}}\big) |\vx{x}|^2 
+ 2\alpha_\text{d} \vx{x}^\top \vx{P} \vx{d}_{\mu}^{\perp}.
\end{equation}

\noindent Hence, $V(\mathbf{x})$ in~\eqref{eq:Vcandid} is a Lyapunov function for the cross-track error dynamics, which satisfies 
$\tfrac{1}{2} \lambda_{\text{min}}(\mathbf{P}) |\mathbf{x}|^2 \leq V(\mathbf{x}) \leq \tfrac{1}{2} \lambda_{\text{max}}(\mathbf{P}) |\mathbf{x}|^2$ for 
all $\mathbf{x}$. Additionally, since $V(\mathbf{x})$ is radially unbounded (i.e., $V(\mathbf{x})\to\infty$ as $|\mathbf{x}| \to \infty$), when $\vx{d}_\mu^{\perp}=\mathbf{0}$, the origin $\mathbf{x}=\mathbf{0}$ is 
globally exponentially stable (GES) for the cross-track error dynamics (see Theorem~4.10 in~\cite{Khalil-2002}).  Furthermore, since $[\varepsilon,\, s]^\top\to \mathbf{0}$ as $t \to \infty$, we have  
$\Delta \theta = 0$. Therefore, on $\mathcal{P}$ where $\varepsilon=0$, it can be seen that  
\begin{equation}\label{eq:ssVel}
\vx{v}^{\tx{d}}(\mathbf{p},\, 0)\Big|_{\mathbf{p}\in \mathcal{P}} = \alpha_{\tx{d}}\mathbf{R}_{\theta_\tx{r}} \begin{bmatrix} \Delta_\tx{LOS} \\ 0 \end{bmatrix} = 
\alpha_{\tx{d}}\Delta_\tx{LOS}\hat{\mathbf{e}}_{_{\theta_\text{r}}}. 
\end{equation}
Consequently, using~\eqref{eq:ssVel} in~\eqref{eq:closed1}  when $d_\mu^\parallel=d_\mu^\perp=0$,  it can be seen that $\dot{\vx{p}}(t) \to 
e_{11}\alpha_{\tx{d}}\Delta_{\tx{LOS}}\hat{\mathbf{e}}_{_{\theta_\text{r}}}$ as $[\varepsilon,\, s]^\top\to \mathbf{0}$ when $t\to \infty$. Since the derivative of the quadratic Lyapunov function $V(\mathbf{x})$ in~\eqref{eq:Vcandid} along the trajectories of 
the cross-track error dynamics  satisfies 
$\dot{V}(x) \leq -\big(\lambda_{\text{min}}(\vx{\Gamma}) - \tfrac{p_{12} +  \sqrt{p_{12}^2+p_{22}^2}}{\Delta_{\tx{LOS}}} -  \sqrt{p_{12}^2+p_{22}^2}\alpha_\text{d} \big) |\vx{x}|^2 , \, \text{ for all } |\mathbf{x}| \geq \| \vx{d}_\mu^{\perp} \|_\infty.$
Therefore, if $\lambda_{\text{min}}(\vx{\Gamma}) > \tfrac{p_{12} }{\Delta_{\text{LOS}}} + \sqrt{p_{12}^2+p_{22}^2}(\alpha_\text{d} + \tfrac{1}{\Delta_{\text{LOS}}})$, the closed-loop dynamics are ISS and convergence to 
$\mathcal{B}_{ d^\ast\lambda_{\text{max}}(\mathbf{P})/\lambda_{\text{min}}(\mathbf{P})}(\mathbf{0})$ holds (see Theorem~4.19 in in~\cite{Khalil-2002}). \hfill$\blacksquare$

The following corollary follows from Proposition~\ref{prop:los}. 
\begin{corollary}\label{corr:invariance}
	Assume the conditions in Proposition~\ref{prop:los}. Then, under $\mathbf{d}_\mu=\mathbf{0}$, the set $\mathcal{P}\times \{s=0\}$ is an invariant set for the microswimmer closed-loop dynamics under the control law ~\eqref{eq:nomCont} with  $\hat{e}_{11}=1$, $\hat{\mathbf{d}}_\mu=\mathbf{0}$, and the ILOS guidance law $\vx{v}^{\text{d}}=\vx{v}^{\text{d}}(\mathbf{p},s)$ 
	given by~\eqref{eq:los} and~\eqref{eq:losdynvar}. Furthermore, once the 
	path $\mathcal{P}$ in~\eqref{eq:line} is made invariant, the speed of the microswimmer along $\mathcal{P}$
	is
	\begin{equation}\label{eq:speedSS}
	v^{\star} = e_{11}\alpha_{\tx{d}}\Delta_{\tx{LOS}}.  
	\end{equation}
\end{corollary}
%

It is possible to further simplify the conditions in~\eqref{eq:stabCond} in a way that 
we obtain inequality constraints on the design parameters $k_\tx{d}$, $\alpha_\tx{d}$, 
$\sigma_0$, and $\Delta_{\text{LOS}}$. \textcolor{black}{First, it can be shown that the roots $r_i,\; i=1,2$, of the 
	quadratic polynomial $P(\lambda) = \lambda^2 - \big(\frac{k_\tx{d}}{\alpha_\tx{d}}+ \frac{p_{11}}{p_{12}} + 
	\frac{\sigma_0 p_{22}}{p_{12}}\big)\lambda+
	\frac{\sigma_0 p_{11} p_{22}}{p_{12}^2}-\frac{1}{4} \Big( \frac{\sigma_0 p_{11}}{p_{12}} 
	+  (1+\frac{k_\tx{d}}{\alpha_\tx{d}})\frac{p_{22}}{p_{12}}    \Big)^2
	$ satisfy the relationship $\lambda_{i\Gamma} = 2\alpha_\tx{d}p_{12} r_i$, $i=1,\; 2$ with $\lambda_{i\Gamma}$ being the 
	eigenvalues of $\Gamma$  in~\eqref{eq:stabCond}.} Next, by computing the 
eigenvalues of $\Gamma$ via finding the roots of $P(\cdot)$ and under the simplifying assumption $p_{11}=p_{22}$, it can be shown that the two inequalities 
\begin{equation}\label{eq:stabCond2}
\frac{k_\tx{d}}{\alpha_\tx{d}}\geq 0,\; 
(\sigma_0 \alpha_\tx{d} \Delta_{\text{LOS}})^2 + 2\alpha_\tx{d}\Delta_{\text{LOS}}(1+\frac{k_\tx{d}}{\alpha_\tx{d}})\leq 1
\end{equation}
guarantee that the conditions in~\eqref{eq:stabCond} are satisfied.

\subsection{Step 2: ODS-based quadratic program}
\label{subsec:ods}

In this section, we present a feedback control solution for the microrobot 
LFC problem based on the optimal decision strategy (ODS) framework~\cite{spong1986control}. Considering the ILOS-based reference 
vector field in~\eqref{eq:los} and~\eqref{eq:losdynvar}, it is clear from Proposition~\ref{prop:los} that 
if the control input $\mathbf{u}$ is designed such that the closed-loop dynamics 
are driven by the ILOS-based guidance law, then the 
LFC problem objective is achieved. However, the step-out frequency limitation 
\begin{equation}\label{eq:ineq}
|\mathbf{u}(t)| \leq \Omega_{\tx{SO}}, 
\end{equation}
on the rotational frequency of the helical microswimmer about its 
axis constrains the magnitude of the angular velocity command input $\mathbf{u}$. Using the ODS 
framework, we will address this constraint. 

\begin{figure}[t]
	\centering
	\includegraphics[scale=0.30]{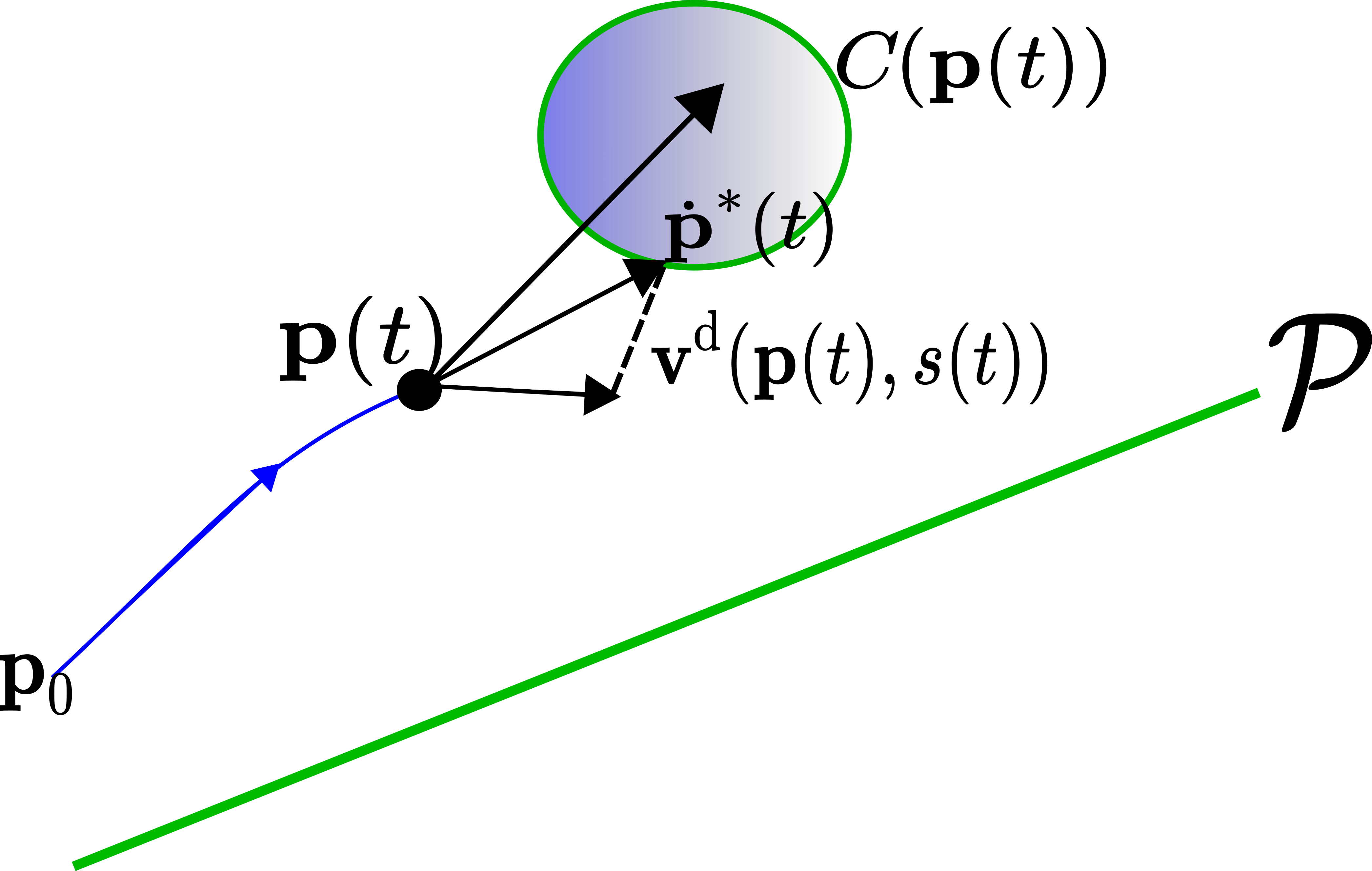}
	\caption{Geometric interpretation of optimal decision strategy.}
	\label{fig:odsFramework}
\end{figure}

ODS-based control is a pointwise optimal control solution
that, in the context of our problem, minimizes the deviation between the vector field of the microswimmer open-loop dynamics 
in~\eqref{eq:eq_dyn} and the ILOS-based reference vector field in~\eqref{eq:los} and~\eqref{eq:losdynvar}, while 
respecting the step-out frequency constraint in~\eqref{eq:ineq}. In order to state the ODS-based control scheme 
for the microswimmer, let $\mathbf{p}(t,t_0,\mathbf{p}_0,\mathbf{u}(t))$, or $\mathbf{p}(t,\mathbf{u}(t))$ or 
$\mathbf{p}(t)$ for short, denote the solution to~\eqref{eq:eq_dyn} corresponding to the control input 
$t\mapsto \mathbf{u}(t)$ and initial position $\mathbf{p}_0$ at time $t_0$. For each solution $\mathbf{p}(t)$, we 
define the set of permissible velocity vectors $C_t$ to be the translation by $\mathbf{p}(t)$ of the set
\begin{equation}\label{eq:permissible}
C(\mathbf{p}(t)) := \big\{ \mathbf{v}(t,\mathbf{w}) \in \mathbb{R}^2 \big| \mathbf{v} = \mathbf{d}_\mu + e_{11}\mathbf{w},\; 
|\mathbf{w}|\leq \Omega_{\text{SO}} \big\}. 
\end{equation}
Therefore, for any control input $\mathbf{u}(t)$ that respects the step-out frequency constraint in~\eqref{eq:ineq}, 
the velocity of the microswimmer lies in the set $C(\mathbf{p}(t))$ (see Figure~\ref{fig:odsFramework}). 

In the case of the microrobot LFC, we choose the ODS-based control law $\mathbf{u}(t)$ in a way that at each time $t$, the instantaneous 
velocity of the microswimmer is ``nearest'' to the ILOS-based guidance law $\mathbf{v}^{\tx{d}}\big(\mathbf{p}(t),s(t)\big)$  
given by~\eqref{eq:los},~\eqref{eq:losdynvar}
in the norm on $\mathbb{R}^2$ defined by some positive definite matrix $\mathbf{Q}$. In other words, the ODS-based
control law at each $t$ is the minimizing solution to 
\begin{equation}\label{eq:permissible2}
\begin{aligned}
\underset{\mathbf{v}\in C(\mathbf{p}(t))}{\text{min.}} \;  \Big\{  \big[\mathbf{v} - \mathbf{v}^{\tx{d}}\big(\mathbf{p}(t),s(t)\big)  \big]^{\top} \mathbf{Q}   \big[ \mathbf{v} - \mathbf{v}^{\tx{d}}\big(\mathbf{p}(t),s(t)\big) \big] \Big\}.  
\end{aligned}
\end{equation}
The minimization problem in~\eqref{eq:permissible2}
is equivalent to the explicit minimization in $\mathbf{u}(t)$, 
\begin{equation}\label{eq:tempODS}
\begin{aligned}
\underset{\mathbf{u}}{\text{min.}} \; & \Big\{ \big[ e_{11}\vx{u}(t) + {\vx{d}}_{\mu}(t) -  \mathbf{v}^{\tx{d}}(\mathbf{p}(t),s(t)) \big]^{\top} \mathbf{Q} \\
&  \big[ e_{11}\vx{u}(t) + {\vx{d}}_{\mu}(t) -  \mathbf{v}^{\tx{d}}(\mathbf{p}(t),s(t)) \big] \Big\} \\
\text{subject to} & \;\;\; |\mathbf{u}(t)| \leq \Omega_{\text{SO}}, 
\end{aligned}
\end{equation}
which, in turn, can be shown to be equivalent to the QP
\begin{equation}\label{eq:ODS}
\begin{aligned}
\underset{\vx{u}}{\text{min.}} \; & \big\{ \frac{1}{2} \vx{u}^{\top} \mathbf{A}_{\mu} \vx{u} + \vx{G}^{\top}_{\mu}(\vx{p},s) \vx{u} \big\} \\
\text{subject to} & \;\;\; \vx{u}^{\top} \vx{u} \leq \Omega_{\text{SO}}^2, 
\end{aligned}
\end{equation}
where
\begin{equation}\label{eq:ODS2}
\textcolor{black}{
	\begin{aligned}
	\vx{A}_{\mu} := e_{11}^2 \vx{Q}, \\
	\vx{G}_{\mu}(\vx{p},s) := \frac{-\vx{A}_{\mu}}{e_{11}} \big( \vx{d}_{\mu} -  \vx{v}^{\tx{d}}(\vx{p},s) \big), 
	\end{aligned}
}
\end{equation}

The QP in~\eqref{eq:ODS}, which involves minimizing a quadratic over a sphere, is called a trust region subproblem (TRS)~\cite{hager2001minimizing}.  It is well-known from the TRS literature (see, e.g., Theorem 1.1 
in~\cite{adachi2017solving}) that $\vx{u}$ is a 
solution to the TRS in~\eqref{eq:ODS} \emph{if and only if} there exists $\lambda^\ast\geq 0$ such that 
\begin{subequations}\label{eq:TRSiff2}
	\begin{align}
		| \vx{u}^\ast | \leq \Omega_{\text{SO}}, 
		\label{eq:a}
		\\
		\big(\vx{A}_{\mu} + \lambda^\ast \vx{I}\big)  \vx{u}^\ast = -\vx{G}_{\mu}(\vx{p},s),
		\label{eq:b}\\
		\lambda^\ast (\Omega_{\text{SO}} - | \vx{u}^\ast |) = 0, 
		\label{eq:c}
		\\
		\vx{A}_{\mu} + \lambda^\ast \vx{I} \succeq \vx{0}.
		\label{eq:d}
	\end{align}
\end{subequations}
Using proper state transformations, it is possible to restate 
the ODS problem in~\eqref{eq:tempODS} such 
that  $\vx{Q}$ becomes diagonal.  In the rest of this paper and without loss of 
generality, we assume that $\vx{Q}$ in the ODS problem in~\eqref{eq:tempODS} has 
the diagonal form $\vx{Q}= \text{diag}\{q_1,\, q_2\}$, where $q_1>0$ and $q_2>0$ are design parameters.  
The following proposition provides the solutions to the TRS in~\eqref{eq:ODS} and~\eqref{eq:ODS2}. 
\begin{proposition}\label{prop:TRSsol}
	Consider the QP in~\eqref{eq:ODS} with the matrix $\mathbf{Q}= \text{diag}\{q_1,\, q_2\}$. Then, 
	the optimal pointwise angular velocity control input, which solves~\eqref{eq:ODS}, is 
	given by  	
	\begin{equation}\label{eq:rawSol}
	\vx{u}^{\ast}(\vx{p},s) =
	\begin{cases}
	-\vx{A}^{-1}_{\mu}\vx{G}_{\mu}(\vx{p},s) & \text{ if } \; | \vx{u}^\ast | < \Omega_{\text{SO}}\\
	-\big(\vx{A}_{\mu} + \lambda^\ast \vx{I}\big)^{-1}  \vx{G}_{\mu}(\vx{p},s) & \text{ if } \; | \vx{u}^\ast | = \Omega_{\text{SO}}
	\end{cases}       
	\end{equation}
	where $\vx{A}_{\mu}$ and $\vx{G}_{\mu}(\vx{p},s)=[G_{1\mu}(\vx{p},s),\, G_{2\mu}(\vx{p},s)]^\top$  are defined in~\eqref{eq:ODS2}, and $\lambda^\ast$ is a non-negative root to the following algebraic equation
	\begin{equation}\label{eq:algSol}
	\frac{G_{1\mu}^2(\vx{p},s)}{\big(e_{11}^{\textcolor{black}{2}}q_1+\lambda^{\ast}\big)^2} + 
	\frac{G_{2\mu}^2(\vx{p},s)}{\big(e_{11}^{\textcolor{black}{2}}q_2+\lambda^{\ast}\big)^2} = \Omega_{\text{SO}}^2.
	\end{equation}
	Furthermore, the angular velocity control input $\vx{u}^{\ast}(\vx{p})$ will be absolutely continuous if and only if $q_1=q_2=\tfrac{1}{\Omega_0 e_{11}^2}$ for some $\Omega_0>0$,  and given by 
	\begin{equation}\label{eq:solFinal}
	\vx{u}^{\ast}(\vx{p},s) =
	\begin{cases}
	-\Omega_{0} \vx{G}_{\mu}(\vx{p},s)  & \text{ if } \; | \vx{G}_{\mu}(\vx{p},s) | < \frac{\Omega_{SO}}{\Omega_{0}}\\
	-\Omega_{\text{SO}}\frac{\vx{G}_{\mu}(\vx{p},s)}{|\vx{G}_{\mu}(\vx{p},s)|} & \text{ if } 
	\; | \vx{G}_{\mu}(\vx{p},s) | \geq \frac{\Omega_{SO}}{\Omega_{0}}
	\end{cases}       
	\end{equation}
\end{proposition}

\textbf{Proof.} Consider the necessary and sufficient conditions in~\eqref{eq:TRSiff2}. Suppose that $| \vx{u}^\ast | < \Omega_{\text{SO}}$. 
Then, according to \eqref{eq:c}, $\lambda^\ast=0$. Hence, under $| \vx{u}^\ast | < \Omega_{\text{SO}}$ and from~\eqref{eq:b}, it can be seen that $\vx{u}^{\ast}(\vx{p},s)=-\vx{A}^{-1}_{\mu}\vx{G}_{\mu}(\vx{p},s)$. 
Next, suppose $| \vx{u}^\ast | = \Omega_{\text{SO}}$. From~\eqref{eq:b}, it follows that 
\begin{equation}\label{eq:temp11}
\vx{u}^\ast = -\big(\vx{A}_{\mu} + \lambda^\ast \vx{I}\big)^{-1}  \vx{G}_{\mu}(\vx{p},s). 
\end{equation}
In order to find the constant $\lambda^\ast$ in the previous equation, we use~\eqref{eq:temp11} to get
\begin{equation}\label{eq:temp22}
\vx{u}^{\ast\top} \vx{u}^\ast = \vx{G}_{\mu}(\vx{p},s)^\top \big(\vx{A}_{\mu} + \lambda^\ast \vx{I}\big)^{-2}  \vx{G}_{\mu}(\vx{p},s). 
\end{equation}
Since 
\begin{equation}\label{eq:temp33}
(\vx{A}_{\mu} + \lambda^\ast \vx{I})^{-1} \vx{G}_{\mu}(\vx{p},s) = \begin{bmatrix}  \frac{G_{1\mu}(\vx{p},s)}{e_{11}^{\textcolor{black}{2}}q_1 + \lambda^\ast} \\ 
\frac{G_{2\mu}(\vx{p},s)}{e_{11}^{\textcolor{black}{2}}q_2 + \lambda^\ast} \end{bmatrix},  
\end{equation}
the right hand side of the equation in~\eqref{eq:temp22} is equal to $\tfrac{G_{1\mu}^2(\vx{p},s)}{\big(e_{11}^{\textcolor{black}{2}}q_1+\lambda^{\ast}\big)^2} + 
\tfrac{G_{2\mu}^2(\vx{p},s)}{\big(e_{11}^{\textcolor{black}{2}}q_2+\lambda^{\ast}\big)^2}$. Therefore, since $\vx{u}^{\ast\top} \vx{u}^\ast=\Omega_{\text{SO}}^2$, the constant $\lambda^\ast$ satisfies~\eqref{eq:temp22} if the algebraic 
equation in~\eqref{eq:algSol} holds. Next, we define 
\begin{equation}\label{eq:tempn1}
\tilde{\lambda} := \lambda^\ast + \Omega_1, \tilde{\Omega} := \Omega_2 - \Omega_1, \Omega_1 := q_1 e_{11}^{\textcolor{black}{2}}, \text{ and }  \Omega_2 := q_2 e_{11}^{\textcolor{black}{2}}.
\end{equation}
Using the definitions in~\eqref{eq:tempn1}, the algebraic equation in~\eqref{eq:algSol} can be transformed 
into the following equivalent quartic equation
\begin{equation}\label{eq:tempn2}
\tilde{\lambda}^4 + 2\tilde{\Omega} \tilde{\lambda}^3 - \big[ \tfrac{ |\vx{G}_{\mu}(\vx{p},s)|^2}{\Omega_{\text{SO}}^2} - \tilde{\Omega}^2 \big] \tilde{\lambda}^2 - 2\tilde{\Omega}G_{1\mu}^2(\vx{p},s) \tilde{\lambda} - \tilde{\Omega}^2 = 0.
\end{equation}
Also, using the definitions in~\eqref{eq:tempn1}, the control input given by~\eqref{eq:rawSol} can be further simplified to 
\begin{equation}\label{eq:tempn3}
\vx{u}^{\ast}(\vx{p},s) =
\begin{cases}
-[\frac{G_{1\mu}(\vx{p},s)}{\Omega_1},\; \frac{G_{2\mu}(\vx{p},s)}{\Omega_2}]^\top & \text{ if } \; | \vx{u}^\ast(\vx{p},s)| < \Omega_{\text{SO}}\\
-[\frac{G_{1\mu}(\vx{p},s)}{\tilde{\lambda_0}},\; \frac{G_{2\mu}(\vx{p},s)}{\tilde{\lambda_0}+\tilde{\Omega}}]^\top & \text{ if } \; | \vx{u}^\ast(\vx{p},s) | = \Omega_{\text{SO}}
\end{cases}       
\end{equation}
where $\tilde{\lambda_0}$ is a root of the quartic equation in~\eqref{eq:tempn2}. From a standard continuity argument, it can be deduced that 
$\vx{u}^{\ast}(\cdot)$ is continuous for every $\vx{p}\in\mathbb{R}^2$ if and only if 
\begin{equation}\label{eq:tempn4}
\tilde{\lambda_0} = \Omega_1. 
\end{equation}
Hence the pointwise optimal control input $\vx{u}^{\ast}(\cdot)$ is continuous everywhere if and only if 
\begin{equation}\label{eq:tempn5}
\lambda^\ast = 0. 
\end{equation}
Therefore, $q_1=q_2=\tfrac{1}{\Omega_0e_{11}^{\textcolor{black}{2}}}$, for some $\Omega_0>0$. 
Consequently, the equation in~\eqref{eq:algSol} gets reduced to 
\begin{equation}\label{eq:tempAlg}
G_{1\mu}(\vx{p},s)^2 + G_{2\mu}(\vx{p},s)^2 = (\lambda^\ast + \tfrac{1}{\Omega_0})^2 \Omega_\text{SO}^2,  
\end{equation}
From~\eqref{eq:tempAlg}, it can be seen that 
\begin{equation}\label{eq:tempAlg2}
\tfrac{| \vx{G}_{\mu}(\vx{p},s) |}{\Omega_\text{SO}}  = \lambda^\ast + \tfrac{1}{\Omega_0}.  
\end{equation}
where $\lambda^\ast = 0$ due to~\eqref{eq:tempn5}. Therefore, when $|\vx{u}^{\ast}|=\Omega_{\text{SO}}$, the pointwise optimal control input satisfies 
\begin{equation}\label{eq:temp44}
\vx{u}^{\ast}(\mathbf{p})=(\vx{A}_{\mu} + \lambda^\ast \vx{I})^{-1} \vx{G}_{\mu}(\vx{p},s) = \begin{bmatrix}  \frac{G_{1\mu}(\vx{p},s)}{\tfrac{1}{\Omega_0} + \lambda^\ast} \\ 
\frac{G_{2\mu}(\vx{p},s)}{\tfrac{1}{\Omega_0} + \lambda^\ast} \end{bmatrix}.  
\end{equation}
\textcolor{black}{where $\lambda^\ast = 0$ according to \eqref{eq:tempn5}}. Consequently, the equation in~\eqref{eq:temp44}, when $|\vx{u}^{\ast}|=\Omega_{\text{SO}}$, is equivalent to 
\begin{equation}\label{eq:temp55}
\vx{u}^{\ast}(\mathbf{p},s)=  \frac{\Omega_\text{SO}}{| \vx{G}_{\mu}(\vx{p},s) |} \begin{bmatrix}  G_{1\mu}(\vx{p},s) \\ 
G_{2\mu}(\vx{p},s) \end{bmatrix} = \frac{\Omega_\text{SO}}{| \vx{G}_{\mu}(\vx{p},s) |} \vx{G}_{\mu}(\vx{p},s).   
\end{equation}
Furthermore, if $|\vx{u}^{\ast}|<\Omega_{\text{SO}}$, then the pointwise optimal control input 
is given by  $\vx{u}^{\ast}(\vx{p},s)=-\vx{A}^{-1}_{\mu}\vx{G}_{\mu}(\vx{p},s)$. Under $q_1=q_2=\tfrac{1}{\Omega_0 e_{11}^{\textcolor{black}{2}}}$, 
we have that $\mathbf{A}_\mu = e_{11}^{\textcolor{black}{2}}q_1\mathbf{I} = \tfrac{1}{\Omega_0}\mathbf{I}$. Therefore, if $|\vx{u}^{\ast}|<\Omega_{\text{SO}}$, then $\vx{u}^{\ast}(\vx{p},s)=-\Omega_{0}\vx{G}_{\mu}(\vx{p},s)$. \hfill$\blacksquare$

One of the features of the ODS-based control solution to the microswimmer's LFC problem in~\eqref{eq:solFinal} is that it is an  
absolutely continuous function of the position of the  microswimmer. Indeed, the pointwise optimal control input in~\eqref{eq:solFinal} 
is continuously differentiable for almost all values of the microswimmer's position $\mathbf{p}$ and the dynamic variable $s$.  
Furthermore, it is a globally Lipschitz function of $\mathbf{p}$ and $s$. The block diagram of our ODS-based control scheme is depicted 
in Figure~\ref{fig:los}(b) and our solution to the LFC problem is summarized as follows: 
%


\section{Simulation Results}
\label{sec:sims}
In this section we present numerical simulation results to validate 
the performance of the ODS-based control method described in 
Section~\ref{sec:Sol}. \\
Table~\ref{table:micro1} summarizes the
parameters of the magnetic microswimmer used in our simulations, which are equal to the parameter values  
of the experimental helical swimmer prototype in~\cite{mahoney2011velocity}. 
The helical microrobot in~\cite{mahoney2011velocity}, which \emph{does not have} a 
magnetic head, swims in corn syrup with a viscosity of 
approximately $\eta_{\tx{mod}}=2500 \tx{ cps}$ and a density of $\rho = 1.36 \tx{ g/ml}$. The 
mass of the magnetic swimmer is equal to $8.9 \tx{ milligrams}$.  The viscosity and density of the corn syrup along with the physical dimensions of the helical swimmer satisfy the low-Reynolds-number 
regime condition. 
\begin{table*}[t]
	\caption{}
	{\small
		\centering 
		\makebox[\linewidth]{
			\begin{tabular}{|l | l| l|} 
				\hline  
				\textbf{Symbol} & \textbf{Description} & \textbf{Numerical values in simulations}\\ [0.1ex] 
				\hline 
				$\theta_\tx{h}$ & Helix pitch angle. & 45$^{\circ}$\\ 
				$n_\tx{h}$ & Number of turns of the helix. & 3.5 \\
				$r_\tx{h}$ & Helix radius. & 420 $\mu \tx{m}$\\
				$|\mathbf{d}_\tx{g}|$ & Weight of the microswimmer. & $8.7 \times 10^{-5}$ N\\
				$e_{11}$ & Physical parameter of the microswimmer given by~\eqref{eq:e11}. & 9.3$\times$10\textsuperscript{-5} m\\
				\hline 
			\end{tabular} 
		}
		\label{table:micro1} 
	}
	\begin{center}
		{The physical parameters of the helical swimmer in~\cite{mahoney2011velocity}}. 
	\end{center}
\end{table*} 
We would like the microrobot to converge to  
$\mathcal{P}:=\{ p \in \mathbb{R}^2 : p = \tau\hat{e}_{_{0}} ,\, \tau\in \mathbb{R} \}$,
where $\hat{e}_{_{0}}=[1,\, 0]^\top$, subject to control input 
saturation limit $\Omega_{_\tx{SO}}=2.8\,\tx{Hz}$. In our simulation studies, we have assumed that 
knowledge of neither the weight of the microswimmer nor its environment are available. Hence, in all 
our simulations, we set $\hat{\mathbf{d}}_\mu=\mathbf{0}$ (lack of knowledge about the disturbances) and $\hat{e}_{11}=1$ (lack of knowledge about the fluid environment and swimmer's physical parameters) in our control scheme. We have chosen the look-ahead distance parameter $\Delta_\tx{LOS}$ in the ILOS guidance law 
in~\eqref{eq:los2} to be equal to $0.75 \tx{ mm}$. Furthermore, we have chosen $\alpha_\tx{d}=600$, 
$\sigma_0=0.01$, and $k_\tx{d}=0.15$. In the pointwise optimal control law in~\eqref{eq:solFinal}, 
we have chosen $\Omega_0=1\,\tx{Hz}$. 
In our simulation studies, we have initialized the microrobot 
position at $p(0)=[0\,\tx{mm},\; -40.0\,\tx{mm}]^{\top}$. In addition to 
the proposed ODS-based scheme, where we use the ILOS-based guidance 
law, we also use an ODS-based scheme with the conventional LOS guidance 
law where there are no integral actions  embedded in the guidance law in two 
different cases. In our first simulation study, we used the same parameters 
for the conventional LOS-based guidance law. In order to improve the tracking performance of the 
conventional LOS-based guidance law in the presence of unknown weight and 
microswimmer physical parameters, we increased the value of $\alpha_\text{d}$ in a 
second set of numerical simulations. 

\begin{figure*}
	\centering
	\begin{subfigure}{0.31\textwidth}
		\includegraphics[width=0.95\textwidth]{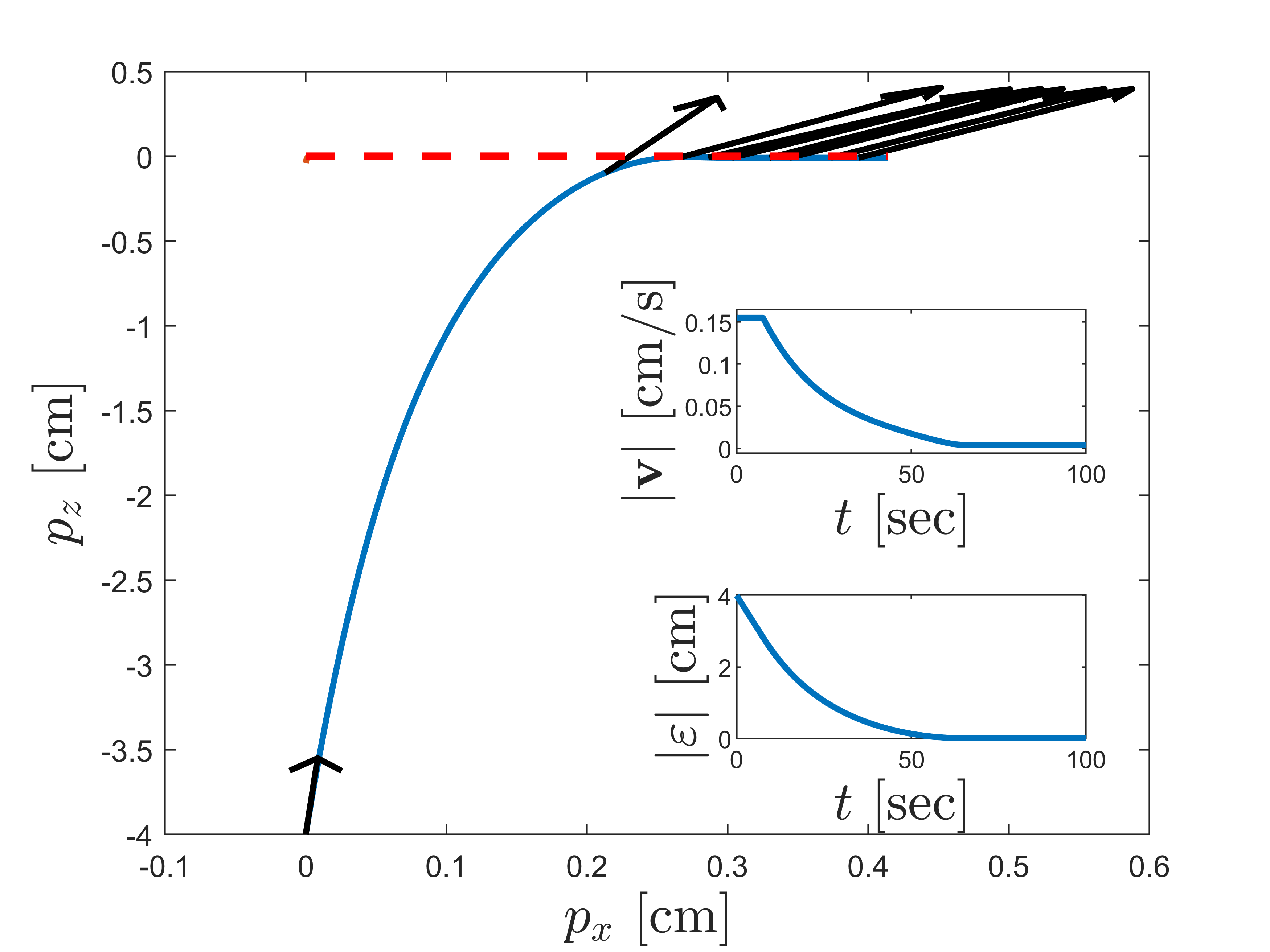} 
		\caption{}
	\end{subfigure}
	\quad
	\begin{subfigure}{0.31\textwidth}
		\includegraphics[width=0.95\textwidth]{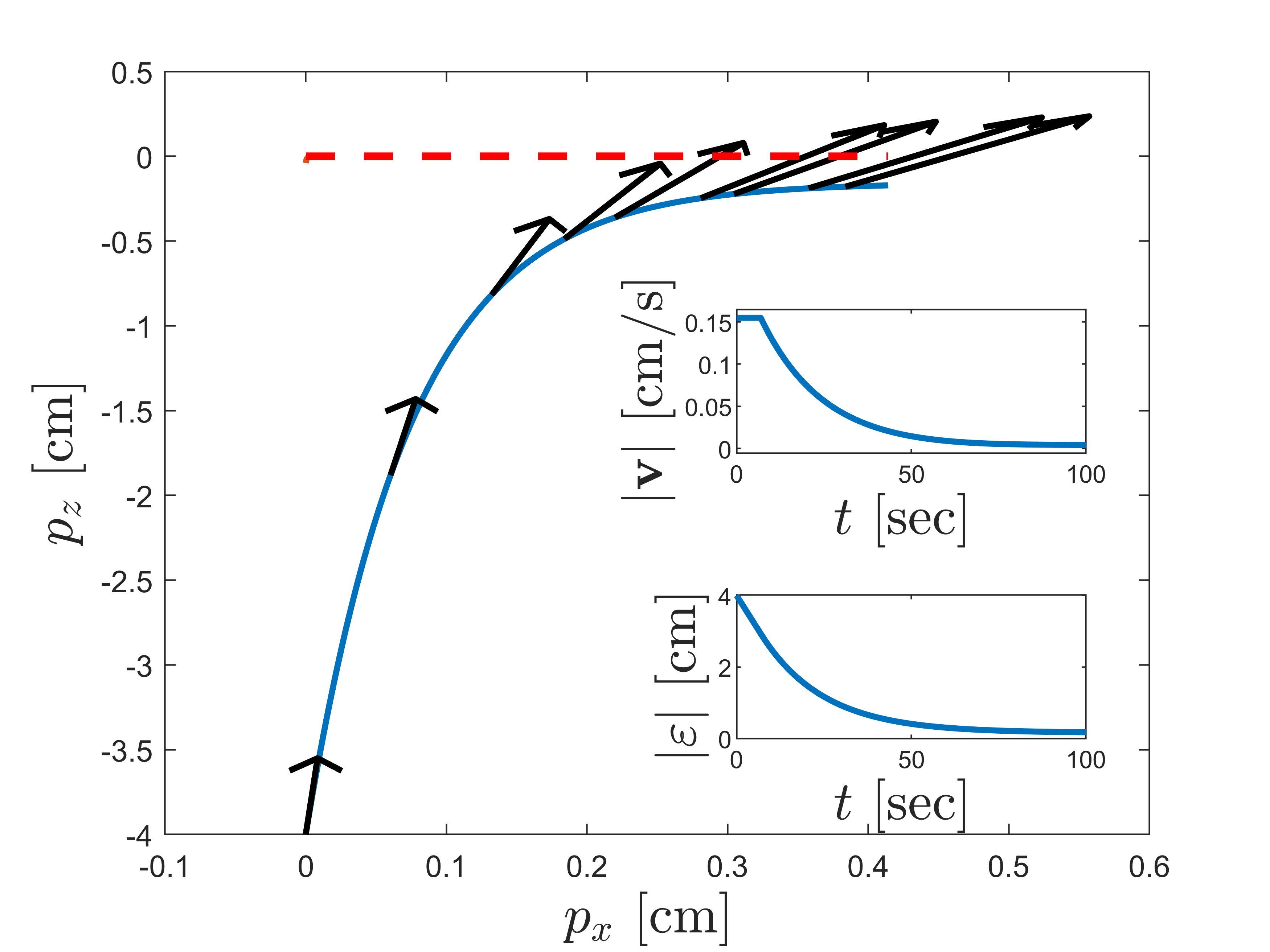} 
		\caption{}
	\end{subfigure}
	\quad
	\begin{subfigure}{0.31\textwidth}
		\includegraphics[width=0.95\textwidth]{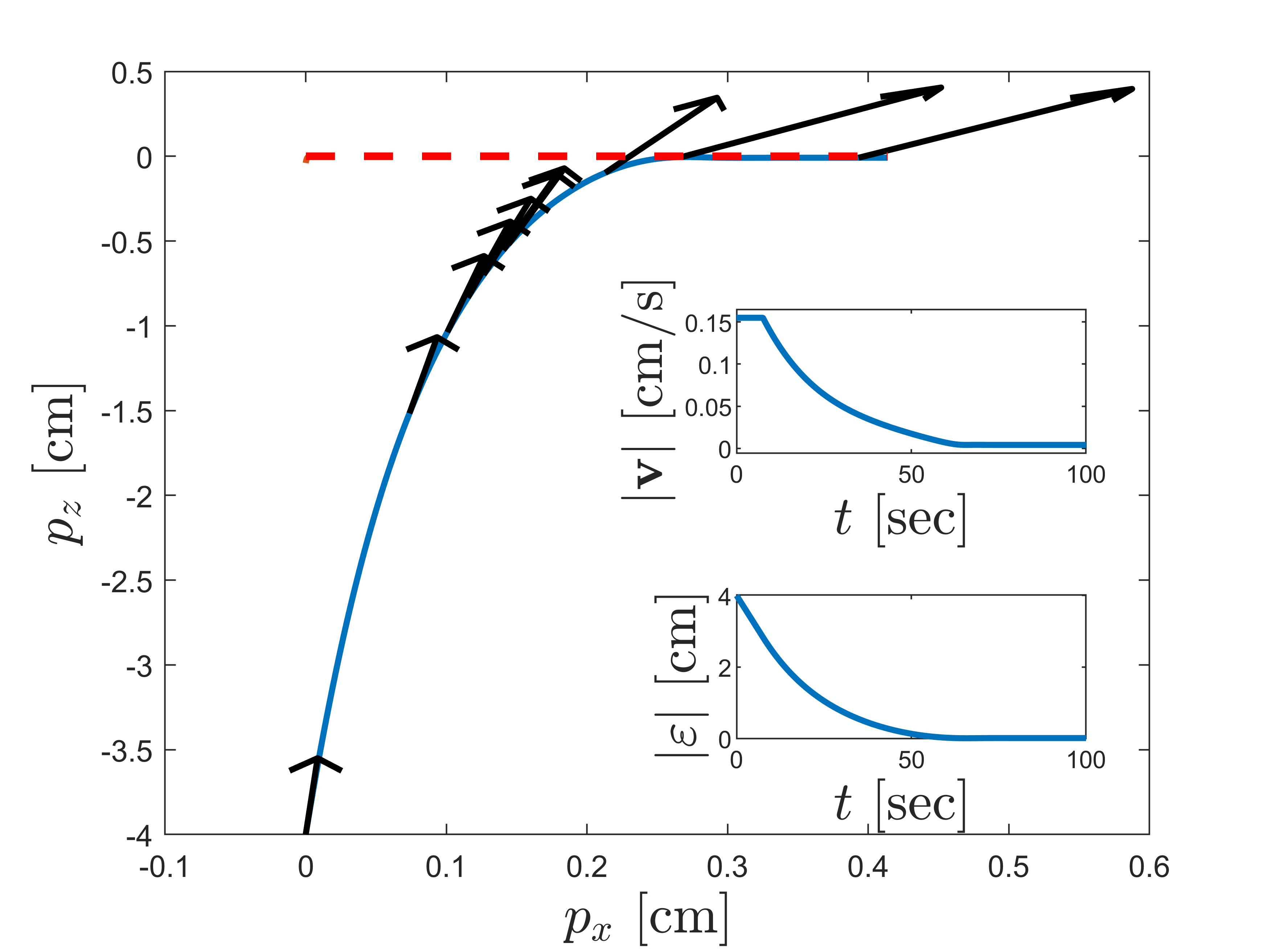} 
		\caption{}
	\end{subfigure}
	\caption{Microrobot position along with cross-tracking error and 
		magnitude of velocity: (a)  the conventional ODS-based scheme in~\cite{mohammadi2018path} with $\alpha_\text{d}=1200$, 
		(b) the conventional ODS-based scheme in~\cite{mohammadi2018path} with $\alpha_\text{d}=600$, 
		and (c) the proposed ODS-based scheme with $\alpha_\text{d}=600$.}
	\label{fig:path_nominal}
\end{figure*}

\begin{figure*}
	\centering
	\begin{subfigure}{0.31\textwidth}
		\includegraphics[width=0.85\textwidth]{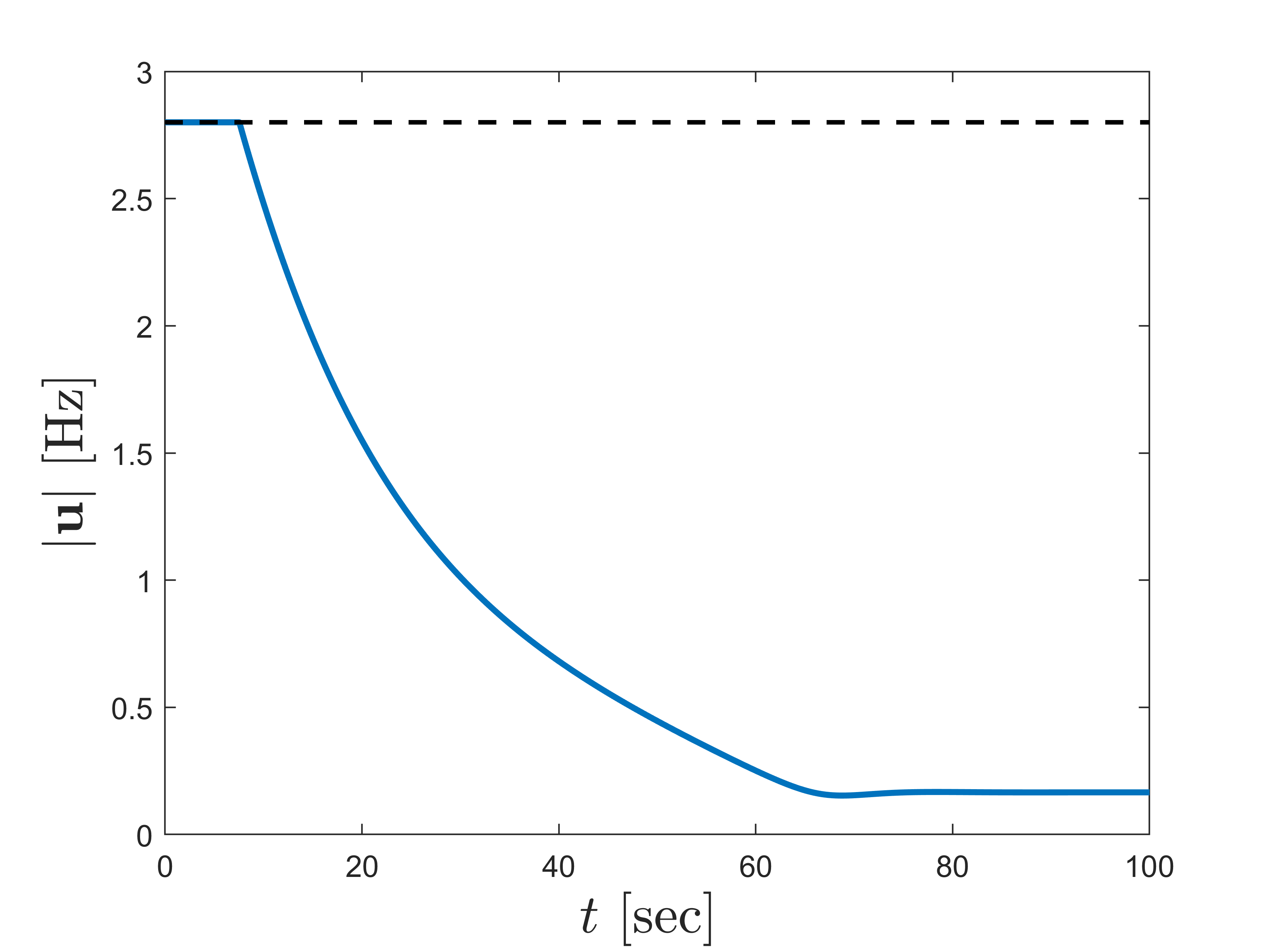} 
		\caption{}
	\end{subfigure}
	\quad
	\begin{subfigure}{0.31\textwidth}
		\includegraphics[width=0.85\textwidth]{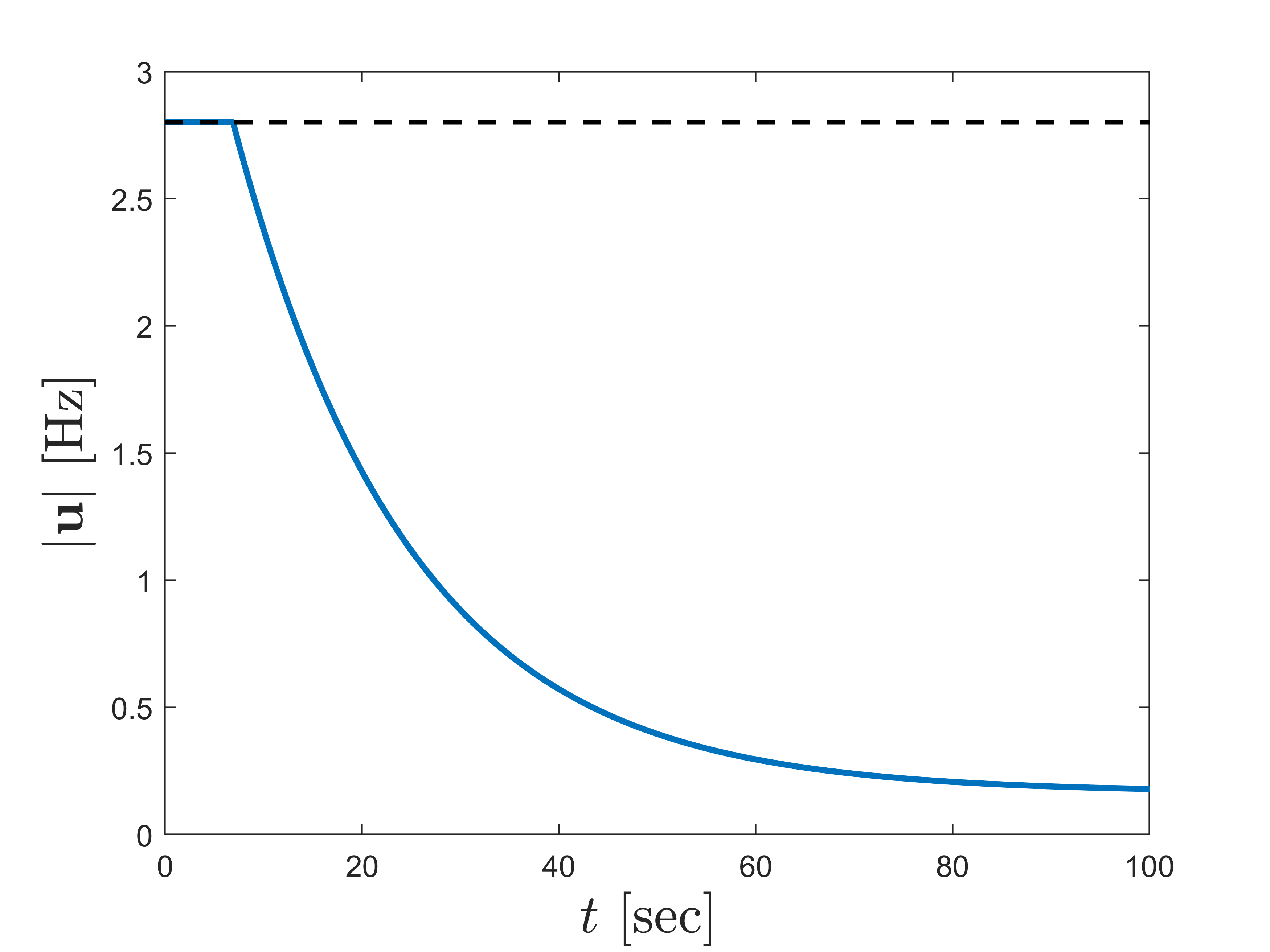} 
		\caption{}
	\end{subfigure}
	\quad
	\begin{subfigure}{0.31\textwidth}
		\includegraphics[width=0.85\textwidth]{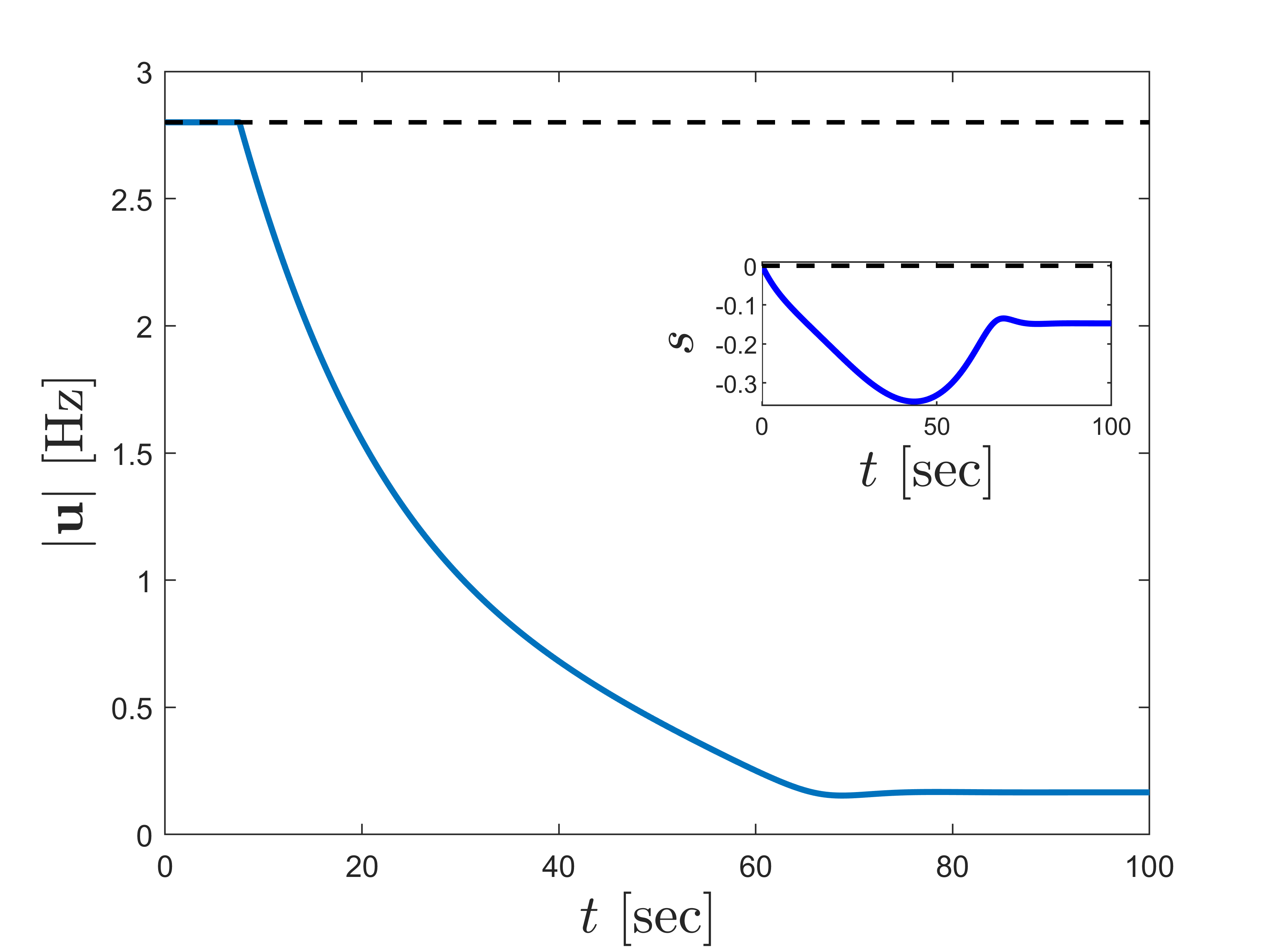} 
		\caption{}
	\end{subfigure}
	\caption{Microrobot control input: (a)  the conventional ODS-based scheme in~\cite{mohammadi2018path} with $\alpha_\text{d}=3000$, 
		(b) the conventional ODS-based scheme in~\cite{mohammadi2018path} with $\alpha_\text{d}=600$, 
		and (c) the proposed ODS-based scheme with $\alpha_\text{d}=600$ and the adaptive variable $s$
		time profile.}
	\label{fig:path_nominal2}
\end{figure*}

Plots in Figure~\ref{fig:path_nominal} depict the path of 
the microrobot under the conventional 
and ILOS-based control inputs that have been generated using the 
ODS-based QP proposed in~\cite{mohammadi2018path}. The inner-plots depict the time profile of the magnitude of the  velocity vector of the microswimmer. Plots in Figure~\ref{fig:path_nominal2} 
depict the control input time profile under the conventional 
and ILOS-based control inputs that have been generated using the 
ODS-based QP. The mean value of the absolute of the cross-track error $|\varepsilon(t)|$ in the last 
$10$ seconds of the simulation, i.e., when $90 \text{ sec}\leq t \leq 100 \text{ sec}$,  were equal to 
$0.79$ mm for the conventional LOS-based guidance law in~\cite{mohammadi2018path} with $\alpha_\tx{d}=1200$, $1.8$ mm for 
the conventional LOS-based guidance law in~\cite{mohammadi2018path} with $\alpha_\tx{d}=600$, and $0.09$ mm for the ILOS-based guidance law with $\alpha_\tx{d}=600$. Although, the tracking error performance 
of the conventional LOS-based guidance law gets improved with increasing the parameter $\alpha_{\tx{d}}$, 
the speed of rotation of the microswimmer remains very high throughout the simulations. In particular, 
whereas the speed of rotation of the microswimmer with the ILOS-based guidance law is approximately 
equal to $0.17 \text{Hz}$ during the steady state, the speed of rotation with the conventional guidance 
law remains higher (around $0.21 \text{Hz}$) during the steady state. 

\textcolor{black}{Discussion on the design parameters: There are four main parameters $k_\tx{d}$, $\alpha_\tx{d}$, $\sigma_0$, and $\Delta_{\text{LOS}}$ that appear in the ILOS-based control law in~\eqref{eq:los2}. As it can be seen from~\eqref{eq:varepsilonDot}, the parameter $\alpha_\tx{d}$ plays a direct role on the cross-track error dynamics. The role of the look-ahead distance parameter $\Delta_{\text{LOS}}$ and the parameter $\sigma_0$ is to simultaneously tune the rate of integration of $s$, which provides us with the integral action, with respect to the cross-track error and to provide a proper geometric direction for the microswimmer as manifested in~\eqref{eq:los2}. The parameter $k_\tx{d}$ adds a further damping action to the dynamic variable $s$ in order to further improving its stability and preventing integrator wind-up. The inequality provided in~\eqref{eq:stabCond2} gives sufficient conditions for Proposition~4.3 to hold. As it can be seen from the chosen parameters in our simulation results $\frac{k_\tx{d}}{\alpha_\tx{d}}\geq 0$ and  $(\sigma_0 \alpha_\tx{d} \Delta_{\text{LOS}})^2 + 2\alpha_\tx{d}\Delta_{\text{LOS}}(1+\frac{k_\tx{d}}{\alpha_\tx{d}})=0.9$.}

\section{Concluding Remarks and Future Research Directions}
\label{sec:conc}
Using the optimal decision strategy (ODS) framework and an integral-line-of-sight 
(ILOS) guidance law, we presented an optimization-based control solution for 
path following control of swimming helical magnetic 
microrobots subject to control input constraints. In addition to formally proving practical convergence 
to desired straight lines with absolutely continuous velocity profiles in the presence of disturbances, 
we also derived closed-form solutions for the pointwise optimal control input that results from the 
trust region subproblem (TRS) arising from the ODS framework for microrobot path following control.  \textcolor{black}{We can think of at least two limitations in our proposed ILOS-based control approach. The first limitation is due to the possibility of integrator wind-up, which is inherent  to the ILOS-based family of controllers. Indeed, the role of the term $(\varepsilon + \sigma_0 s)^2$ in the ILOS-based control law is to decrease the risk of such phenomenon.} \textcolor{black}{The second limitation is due to the physics of the problem and the environment in which the microswimmer might be operating in. In particular, if the disturbances from the microswimmer's ambient environment have impulsive nature such as the systolic blood pressure, they might be able to violate the Hypotheses H1 (synchronous rotation with the external field) and H2 (alignment of the microswimmer central axis with the external field). Under this violation, a more complex dynamical model of the microswimmer such as the one provided in~\cite{mahoney2012non} would be needed for modifying our proposed ILOS-based control law.} 

The proposed methodology leads us to further research avenues 
for controlling helical magnetic microrobots such as way-point tracking 
control in the cluttered areas of the human body,  three-dimensional 
maneuvering control, control in the presence of disturbances such as 
vessel blood flow, and control of a collection of microrobots moving 
in a formation.


\bibliographystyle{ieeetran}        
\bibliography{PhDBiblio}

\end{document}